\documentclass{pasj00}
\usepackage{color}
\draft

\begin{document}
\SetRunningHead{Y.Fukazawa, et al.}{Modeling and Reproducibility of
Suzaku HXD PIN/GSO Background}
\Received{2007/**/**}
\Accepted{2008/**/**}

\title{Modeling and Reproducibility of Suzaku HXD PIN/GSO Background}

\author{Yasushi \textsc{Fukazawa}$^1$, 
Tsunefumi \textsc{Mizuno}$^1$, Shin \textsc{Watanabe}$^2$, \\
Motohide \textsc{Kokubun}$^2$, Hiromitsu \textsc{Takahashi}$^1$, \\
Naomi \textsc{Kawano}$^1$, Sho \textsc{Nishino}$^1$, 
Mahito \textsc{Sasada}$^1$, Hirohisa \textsc{Shirai}$^1$, \\
Takuya \textsc{Takahashi}$^1$, Tomonori \textsc{Yamasaki}$^1$, 
Tomonori \textsc{Yasuda}$^1$, Aya \textsc{Bamba}$^2$, \\
Masanori \textsc{Ohno}$^2$, Tadayuki \textsc{Takahashi}$^2$, 
Masayoshi \textsc{Ushio}$^2$, Teruaki \textsc{Enoto}$^3$, \\
Takao \textsc{Kitaguchi}$^3$, Kazuo \textsc{Makishima}$^{3,4}$, 
Kazuhiro \textsc{Nakazawa}$^3$, Yuichi \textsc{Uehara}$^3$, \\
Shin'ya \textsc{Yamada}$^3$, Takayuki \textsc{Yuasa}$^3$, 
Naoki \textsc{Isobe}$^4$, Madoka \textsc{Kawaharada}$^4$, \\
Takaaki \textsc{Tanaka}$^5$, Makoto \textsc{Tashiro}$^6$, 
Yukikatsu \textsc{Terada}$^6$, and Kazutaka \textsc{Yamaoka}$^7$}
\affil{$^1$Department of Physical Science, Hiroshima University, 1-3-1 Kagamiyama, \\
Higashi-Hiroshima, Hiroshima 739-8526}
\email{fukazawa@hirax7.hepl.hiroshima-u.ac.jp}

\affil{$^2$Department of High Energy Astrophysics, Institute of Space and \\
Astronomical Science (ISAS), 
Japan Aerospace Exploration Agency (JAXA), \\ 3-1-1 Yoshinodai, Sagamihara, Kanagawa 229-8510}

\affil{$^3$Department of Physics, University of Tokyo, 7-3-1 Hongo,
Bunkyo, Tokyo 113-0033}

\affil{$^4$Cosmic Radiation Laboratory, The Institute of Physical and
Chemical Research (RIKEN), \\ 2-1 Hirosawa, Wako, Saitama 351-0198}

\affil{$^5$SLAC National Accelerator Laboratory and KIPAC, Stanford
University, Menlo Park, CA, USA}

\affil{$^6$Department. of Physics, Saitama University, 255 Shimo-Ohkubo,
Saitama-City, Saitama 338-8570}

\affil{$^7$Department of Physics, Aoyama Gakuin University, 5-10-1
Fuchinobe, Sagamihara, Kanagawa 229-8558}


%


\maketitle

\begin{abstract}
Suzaku Hard X-ray Detector (HXD) achieved the lowest background level
than any other previously or currently operational missions sensitive 
in the energy range of 10--600 keV, 
by utilizing PIN photodiodes and GSO scintillators mounted in the BGO
active shields to reject 
particle background and Compton-scattered events as much as possible.
Because it does not have imaging capability nor rocking mode for the 
background monitor,
the sensitivity is limited by the
reproducibility of the non X-ray background (NXB) model.
We modeled the HXD NXB, which varies with time as well as other
satellites with a low-earth orbit, by utilizing several parameters,
including particle monitor counts and satellite orbital/attitude information.
The model background is supplied as an event file in which the 
background events are generated by random numbers, 
and can be analyzed in the same way as the real data.
The reproducibility of the NXB model depends on the event selection 
criteria (such as cut-off rigidity and energy band) 
and the integration time, and the
1$\sigma$ systematic error is estimated to be less than 3\% 
(PIN 15--40 keV) and 1\% (GSO 50--100 keV) for 
more than 10 ksec exposure.
\end{abstract}

\section{Introduction}

The hard X-ray detector (HXD)\citep{takahashi07,kokubun07}
onboard Suzaku \citep{mitsuda07} has been developed to enable observations of astronomical
objects with a good sensitivity in the 10--600 keV band.  
When the HXD data are combined with
the data from the X-ray CCD camera (XIS), they simultaneously cover a
wide energy band, from 0.2 keV to 600 keV.  
The HXD sensor part (HXD-S) consists of 64 Si-PIN photo
diodes and 16 GSO/BGO phoswich counters, shielded by 20 BGO
anti-coincidence counters.  
Such a compound-eye configuration greatly 
reduces the background and the dead-time.  
Each of the phoswich counter
(well counter) units includes 4 Si-PIN photo diodes with 2 mm thickness
and 4 GSO scintillators with 5 mm thickness, located at the bottom of
the BGO well-shaped active collimator.  
The HXD is the first
astronomical detector to utilize such thick Si-PIN diodes and high-Z GSO
scintillators.  
In order to achieve lower background level than any other
previous missions, the HXD has been designed to contain less radio
isotopes -- which may be naturally contained in the detector or 
activated in the orbit -- and reject the background events
very effectively.  
The important goal of the HXD is the achievement of the low
background level by rejecting particle background
and Compton-scattered events as much as possible.
It is found that, considering the detection efficiency,
the residual background level of the HXD is achieved to be 10 mCrab at 
20 keV and 1 Crab at 200 keV, the lowest of all previous missions,
such as BeppoSAX and RXTE.  
Such a low background level enables us to detect
weak sources of several hundreds $\mu$Crab in several tens keV
band or several $m$Crab around 100 keV,
without a necessity for the ``rocking motion'' of the detector for 
simultaneous monitoring of
the background, as has been done with CGRO/OSSE, BeppoSAX/PDS, and RXTE/HEXTE.

Such a low background level and narrow field of view are
unique in order to permit the detection of faint hard sources, give less
source confusion and contribution of the Cosmic X-ray Background, and
provide a rudimentary capability to study the structure for the
emission region of hard
X-rays, especially for supernova remnants, galactic diffuse
X-ray emission, and galaxy clusters.  
Simultaneous observations with the
XIS CCD camera -- with no gap in the spectral coverage -- are very
effective to constrain the time variability of the broad-band spectrum,
and the appreciable effective area up to several hundred keV enables us
to determine the continuum shape accurately, which is especially important
for time-variable Galactic binaries and active galaxies.

Although the HXD achieved the low background level, 
the count rate of the background significantly varies with time, 
depending on mainly two parameters; geomagnetic cut-off rigidity (COR) 
and elapsed time after passages 
of the South Atlantic Anomaly (SAA), but weak dependences 
on other parameters such as earth elevation are found.
Therefore, the NXB must be estimated by taking into account these dependences.
Since the sensitivity is limited by the reproducibility of the NXB, 
a correct NXB model is strongly required.
The HXD team has developed the NXB model with several different methods
and unified them by combining advantages of each method.
Here we report the HXD background modeling and its reproducibility.
Corresponding Suzaku documents are also available on the Suzaku web site
as Suzaku memo\footnote{{\tt http://www.astro.isas.jaxa.jp/suzaku/doc/}} 
2007-01, 2007-02, 2007-09, 2008-01, and 2008-03.
In this paper, some technical terms often appear and therefore we summarize
them in table \ref{acronym}.

\section{Characteristics of HXD NXB}

Here we show the characteristics of the HXD NXB.
We utilize the data during the earth occultation, which are 
almost equivalent to the NXB.
The earth albedo is at most 10 \% of the CXB flux below 50 keV
\citep{imhof76,churazov07,frontera07,sazonov07,churazov08}, and the CXB
flux is 5\% of the PIN NXB and $<$0.5\% of the GSO NXB.

\subsection{Summary of the NXB Components}

\begin{figure}[hpbt]
\vspace*{-3cm}
\centerline{\includegraphics[width=10cm]{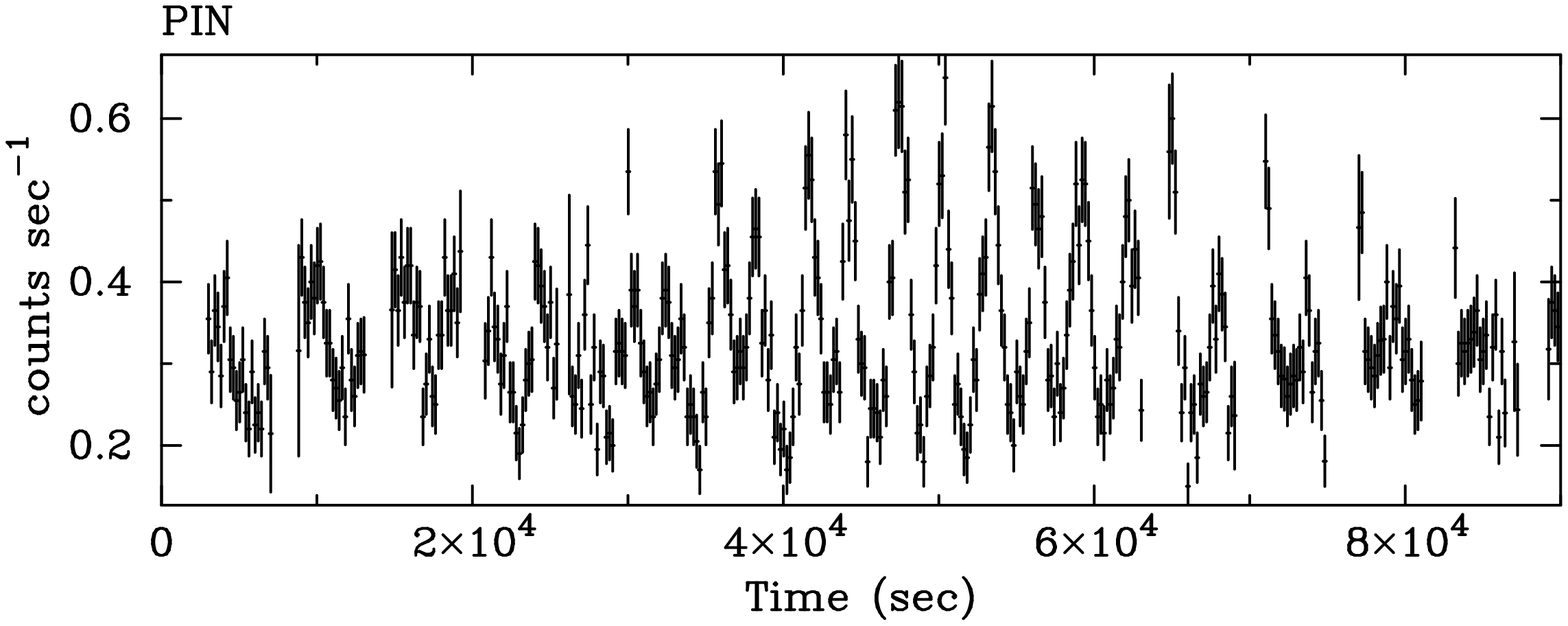}}
\vspace*{-3cm}
\centerline{\includegraphics[width=10cm]{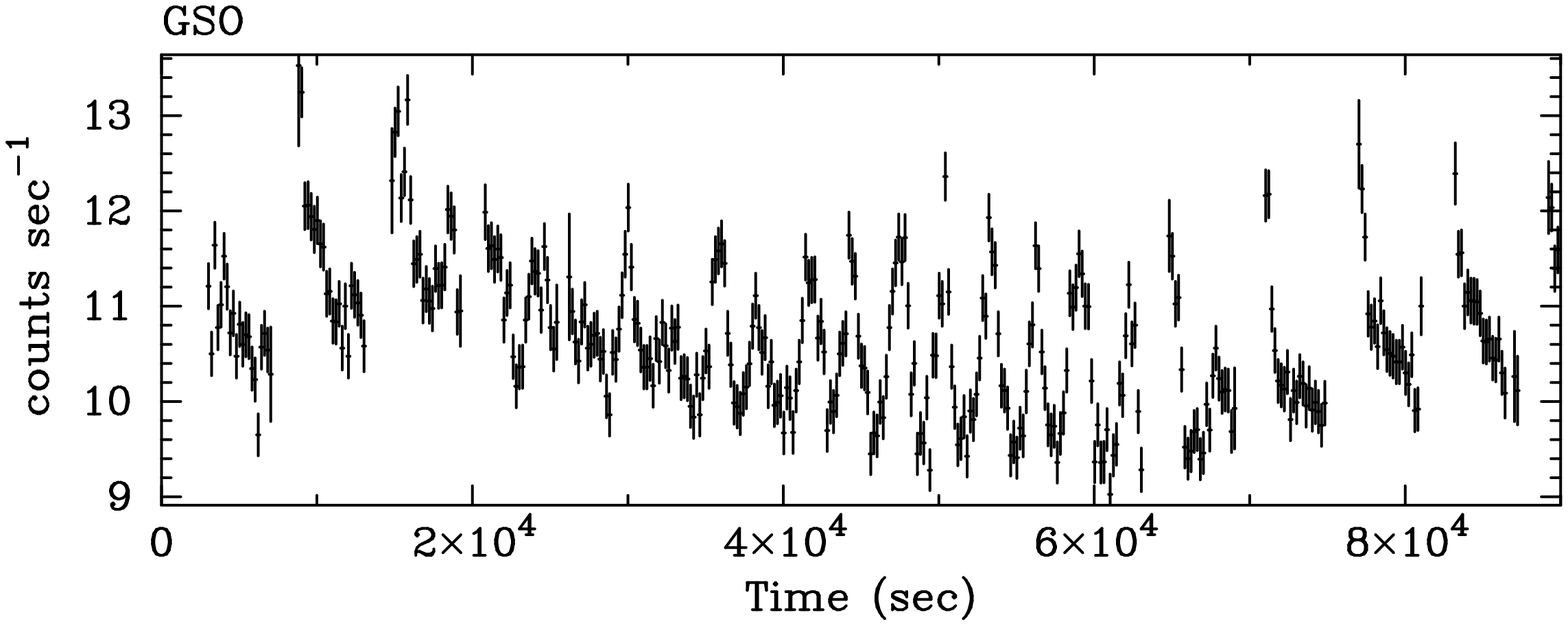}}
\caption{Left and right show examples of light curves of the PIN
 (15--40 keV) and GSO (50--200 keV) background, respectively. 
The variation of COR-dependence is seen at 2$\times10^4$--6$\times10^4$
sec, and exponential-decay components after passages of SAA are seen 
at 0--2$\times10^4$ sec and after 6$\times10^4$ sec.}
\label{bgdlc}
\end{figure}

Detail description of the HXD NXB is written in \citet{kokubun07}.
Here we shortly review it, and present the characteristics which has been
newly found. 
Example of light curves of the HXD background is shown in figure
\ref{bgdlc}.
The HXD background events are mainly caused by natural radio active isotopes,
SAA-induced radio isotopes, primary and secondary cosmic rays, 
and atmospheric albedo neutrons.
Therefore, the count rate of the background significantly varies with time, 
depending on mainly two parameters.
First is the dependence of the COR for primary cosmic rays and
atmospheric neutrons.
This dependence is well tracked by PINUD, which counts the
upper discriminator signal of the PIN at the threshold energy of $\sim$90 keV.
Therefore, the PINUD is insensitive to X-ray events, but 
monitors a flux of real-time charged particles.
Since the PIN diodes are embedded in the thick BGO shields, PINUD counts
the proton above $\sim$100 MeV.
This component is dominant in the variation of PIN NXB and GSO NXB below
140 keV.
The spectra of PIN NXB become harder as the COR decreases 
or the PINUD increases.
The PINUD is considered to be a better indicator of the particle flux 
than the COR, because the particle environment against the COR would
depend on the solar activity.
The second is exponential-decay components after passages of SAA,
due to the activated material bombarded by SAA protons, which is monitored
by the hit count of PINUD integrated during the SAA 
(thus representing the total accumulated dose). 
This component is significant in the GSO band below 100 keV and above 400
keV, while the variation due to this component is smaller for the PIN NXB.
Since there are many activated nuclides with different life times, many time
scales of NXB variation also exist in the range of several seconds to
several hundred days.
These two dependences are the clearest, but other weak dependences have
been found.
For example, the background rate is somewhat dependent on whether the
pointing direction is toward the sky or the earth.
Details are described later in \S\ref{bgddpin}.

\begin{table}[ht]
\caption{Acronym.}
\label{acronym}
\begin{center}
\begin{tabular}{cc}
\hline\hline
HXD & Hard X-ray Detector \\
PIN & p-i-n type Si photo diode in the HXD \\
GSO & GdSiO scintillator in the HXD \\
BGO & BiGdO anti-coincidence scintillator in the HXD \\
BGD & background \\
CXB & Cosmic X-ray background \\
NXB & non X-ray background \\
SAA & South Atlantic Anomaly \\
PINUD & PIN Upper discriminate counter \\
PINUD$_{\rm buildup}$ & convolution of PINUD with an exponential decay
function \\
COR & geomagnetic cut-off rigidity \\
PINUDLCUNIT (bgd\_a or quick) & background model (\S3.1) \\
LCFITDT (bgd\_a or tuned) & background model (\S3.2) \\
LCFIT & background model (\S3.2) \\
GSOHCNT & count rate of GSO event in 450--700 keV \\
GLC & gradually increasing component of the GSO background event in
450--700 keV \\
T\_SAA\_HXD & elapsed time after the SAA \\
\hline
\end{tabular}
\end{center}
\end{table}

\subsection{HXD Operation and NXB}

Apart from the intraday variation, the count rate of the NXB depends
on the HXD operation mode and the activation of detector material with
long life times.
There has been two kinds of important changes of the HXD operation mode.
The lower discrimination level and pulse-shape
discrimination level for the GSO signal determines 
the background rejection efficiency of PIN NXB, and therefore their
level change can cause the change of the PIN NXB.
Due to the noise increase of one PIN possibly caused by the in-orbit 
radiation damage, the bias high voltage to the PIN
was changed twice.
Since the thickness of depletion layer depends on the bias, 
this change slightly affected the energy response of the 16/32 PIN and
thus PIN NXB rate might also be affected.
Therefore, the response matrices corresponding to different bias
voltages were supplied from the HXD team.
The activation due to long-lived radio-active nuclei leads to the
increase of the NXB rate, especially for the GSO.

\begin{figure}[htb]
\vspace*{-3.5cm}
\centerline{\includegraphics[width=0.8\textwidth]{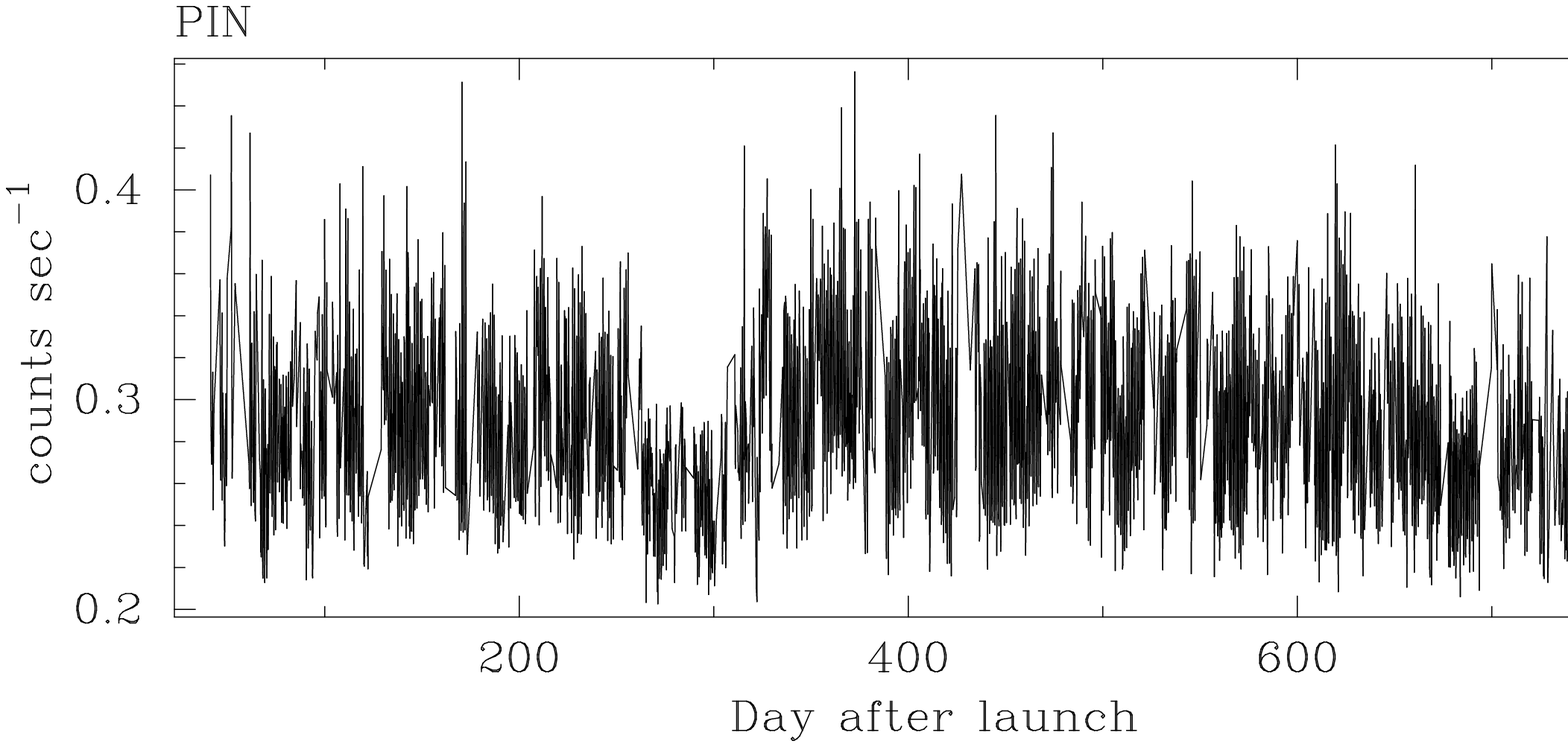}}
\vspace*{-3.5cm}
\centerline{\includegraphics[width=0.8\textwidth]{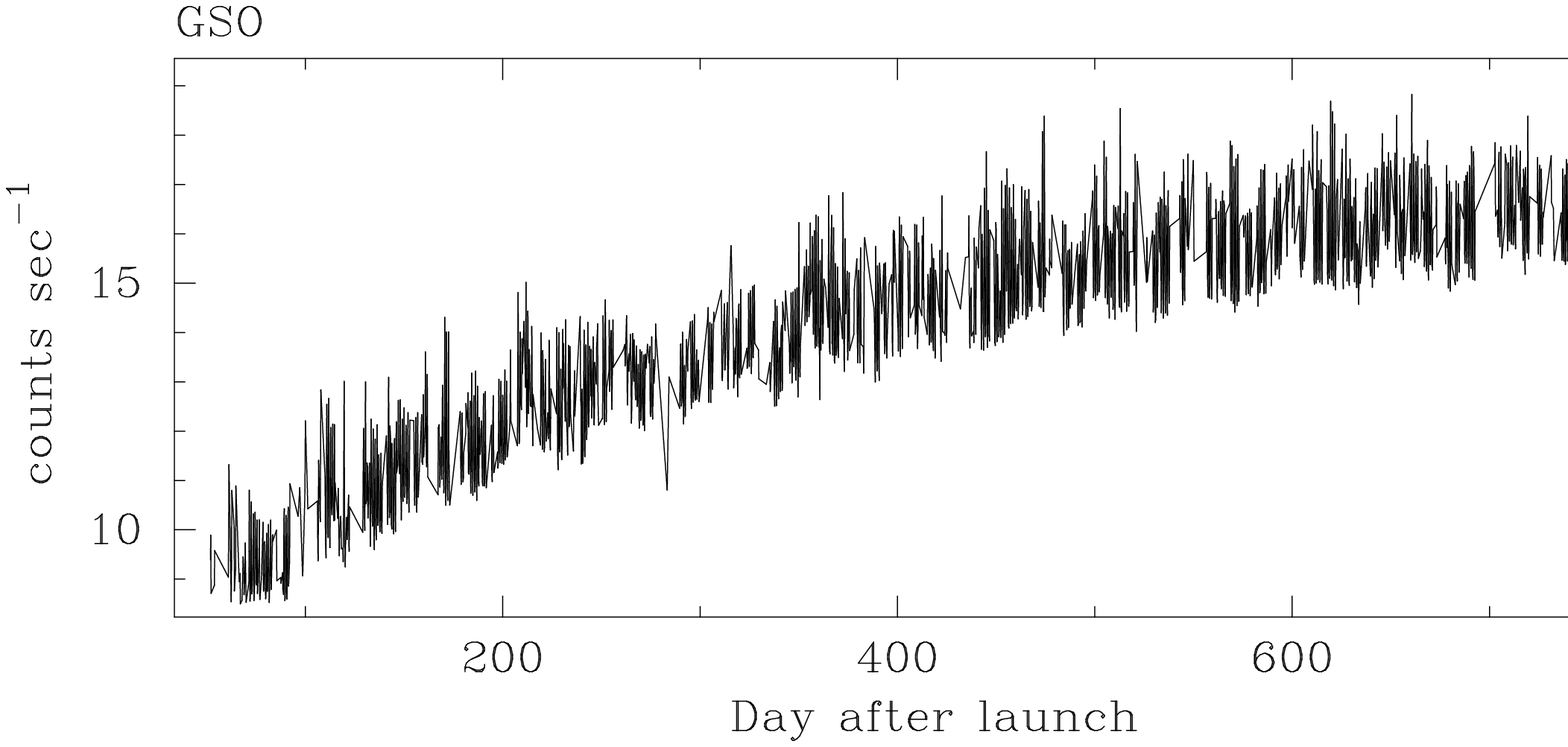}}
\vspace*{1cm}
\caption{Time history of the PIN count rate (15--40 keV) (top) and 
the GSO count rate (50--200 keV) (bottom) of NXB during earth occultation data.
The period of low count rate of the PIN NXB 
corresponds to 2006/03/23 -- 2006/05/13
(day from launch = 256 -- 307).
}
\label{increase}
\end{figure}

Here we show the long term variation of the NXB, concerning the above issues.
Figure \ref{increase} top shows a two-year history of the PIN NXB count rate 
(15--40 keV) during earth occultation data.
The count rate is almost constant, and thus the activation is not
important for the PIN NXB.
However, some discrete or gradual changes
are seen; the most significant are two discrete jumps around a day of
260 and 310.
This is due to the above level changes for the GSO on 2006 March
23 and May 13.
Bias voltages for 16 out of 64 PIN
were reduced from 500 V to 400 V on 2006 May 29 (323 days after the
launch), and those of
additional 16 PIN down to 400 V on 2006 Oct 3 (450 days after the
launch).
Nevertheless, we find no significant changes of the PIN NXB rate 
from this figure.
Apart from those variations, the PIN NXB rate will suffer the increase 
of solar activity in the near future.

\begin{figure}[htb]
\centering
\includegraphics[width=0.4\textwidth]{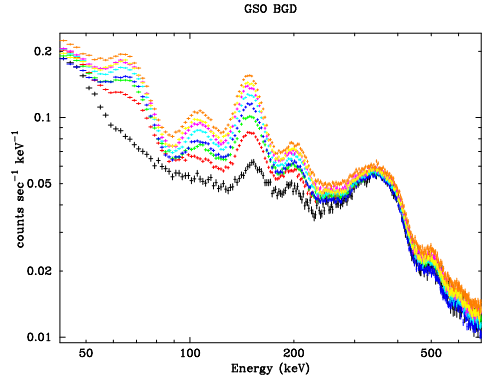}
\caption{
Comparison of the GSO NXB spectra using
earth occultation data in different epochs; 2005-08-31, 2005-12-03, 2006-03-09,
 2006-06-13, 2006-09-26, 2007-01-05, 2007-04-10, and 2007-08-09 from
 bottom to top.
}
\label{GSOincrease}
\end{figure}

In figure \ref{increase} bottom, a two-year history of the GSO count rates 
of the NXB in 50--200 keV during earth occultation data is plotted.
Due to the activation of long-lived nuclides during the SAA passage, 
the count rate gradually increases and is reaching saturation.
The increase is larger in the lower energy band.
Figure \ref{GSOincrease} shows the comparison of spectra of
earth occultation data in different epochs.
Gradual increase of several activation-induced lines are clearly seen.
During 2006 March 14 to May 13 (247--307 days after the
launch), observational mode for the GSO had been
changed several times to optimize the setting parameters before the
guest observer program started.
The GSO count rate below 100 keV, especially around the lower discri
level, was affected by these mode changes.
Although this affection is significant but small, it is not clearly
seen in figure \ref{GSOincrease}.
As a result, the background model should consider these issues.

\section{Modeling of NXB}

The most ideal model of the NXB is based on the first principle; predicting the
NXB count rate by considering the orbital particle environment and the
detector response correctly, for example, by Monte-Carlo simulations.
We in fact predicted the GSO background level due to the activation by
Monte-Carlo simulations \citep{kokubun99,kokubun07} within a factor of 2.
However, in order to reproduce the NXB with a few \% accuracy, it is
more realistic to utilize the actually observed NXB.
In that case, there are usually two types of methods.

One is based on the NXB data base which is sorted by some parameters,
such as COR, and the NXB for the observation is constructed by referring
to parameters to extract the data base.
This method is available when the number of sorting parameters are at
most two; the NXB time variation is not so complex.
Furthermore, it is needed that the background properties, which would
depend on other parameters than sorting parameters, do not change
within the period during which the NXB data base is accumulated.
The background model of the Suzaku XIS applies this method
\citep{tawa08}.
However, strictly speaking, the NXB usually depends on more than two 
parameters, and therefore the accuracy of this method is
somewhat limited.

The second method is to predict the NXB by the empirical function 
which represents the NXB time variation by several parameters.
The function is constructed through the analysis of the real 
time variation of the NXB.
The background models of the Ginga LAC and the RXTE PCA are based 
on this method \citep{hayashida89,jahoda06}.
This method can include many parameters to represent the function, but 
enough statistics of the NXB is necessary to track the time variation of
NXB with a time scale of several hundred seconds.
In the case of low count rate of NXB, it is somewhat difficult to find
modeling parameters.

The count rate of the PIN NXB is not so high
but enough to utilize the second method, while that of the GSO NXB is
high and more variable.
The latter method is more accurate, but cannot prepare the NXB model 
just after the data becomes available, while the former method can
prepare the NXB model quickly.
Therefore, we developed the PIN NXB model using both methods and
the GSO NXB model using only the latter one.
The model background is supplied as an event file by generating the
background event by random numbers, and can be analyzed in the same way
as the real event file.

\subsection{Method based on the data base sorted by PINUD and PINUD
  build-up : {\tt PINUDLCUNIT} (PIN NXB model)}

\begin{figure}[hpbt]
\begin{minipage}[tbhn]{8cm}
\centerline{\includegraphics[width=8cm]{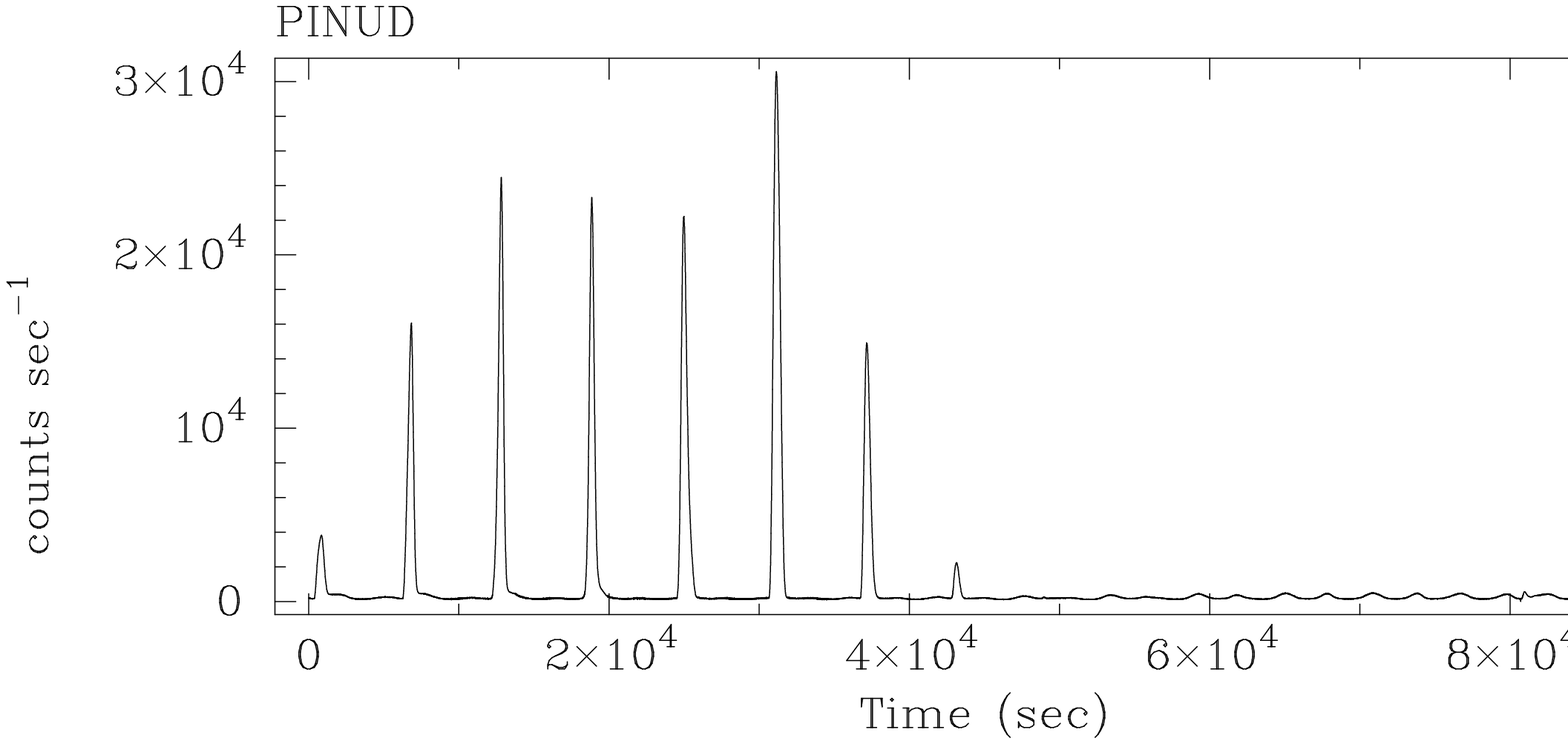}}
\vspace*{-2cm}
\centerline{\includegraphics[width=8cm]{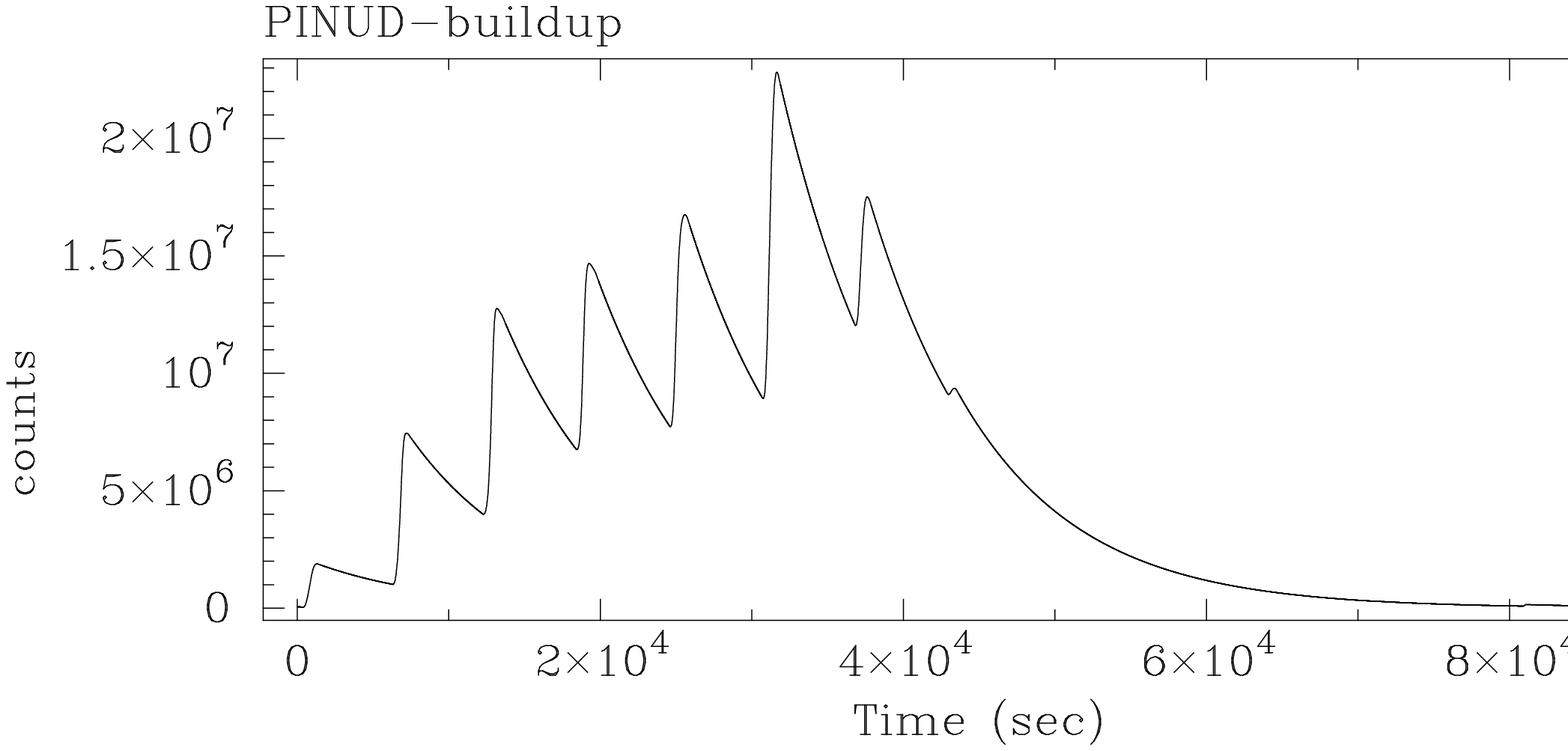}}
\caption{Example of PINUD-buildup. Top panel shows the light curve of
 the PINUD count rate in unit of cts sec$^{-1}$ per one unit (4 PINs), 
and bottom shows the PINUD$_{\rm buildup}$ convolved with
 a time constant of 8000 sec.}
\label{pinudbuildup}
\end{minipage}\quad
\begin{minipage}[tbhn]{8cm}
\centerline{\includegraphics[width=8cm]{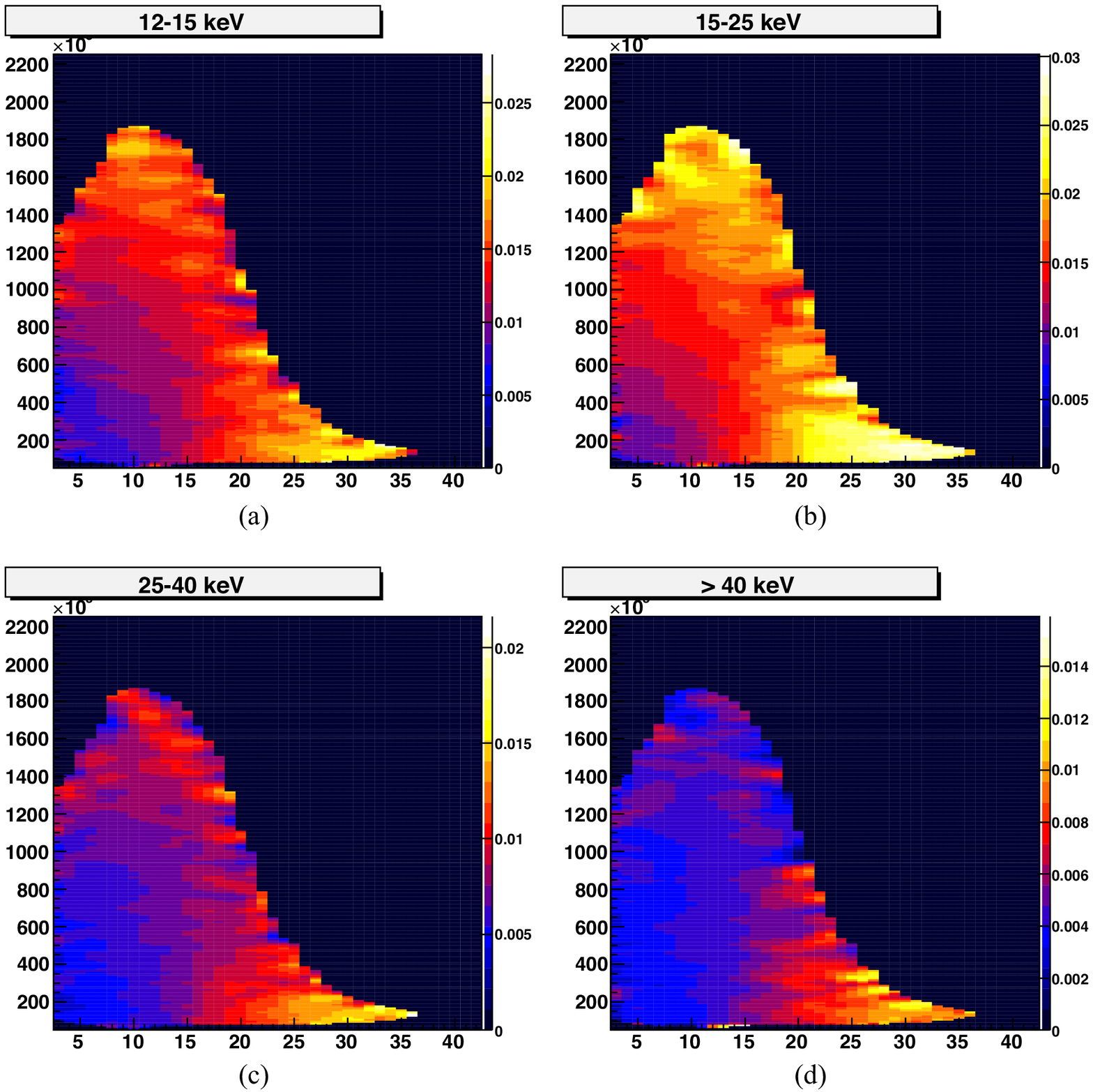}}
\caption{A NXB database of the {\tt PINUDLCUNIT} method 
for (a) 12--15 keV, (b) 15--25
 keV, (c) 25--40 keV, and (d) $>$40 keV. The horizontal and vertical
 axes are PINUD and PINUD$_{\rm buildup}$, respectively. Color
 represents the count rate of the PIN NXB.}
\label{bgdasort}
\end{minipage}
\end{figure}

The method described here utilizes the two-parameter-sorted database, 
and is applied to the PIN NXB model. 
In this method ({\tt PINUDLCUNIT})
\footnote{It is referred as ``quick'' or ``bgd\_a'' in the Suzaku team.
The event FITS files generated with this method are identified by the
keyword of {\tt PINUDLCUNIT} in the {\tt METHOD} record of the FITS file header.}, the NXB model is constructed on the data base of the earth
occultation data.
The real time PINUD count rate can be used as a good indicator of the
NXB component.
Another NXB component related with the activation cannot be modeled by
simply using PINUD.
Therefore, we introduced a parameter ``PINUD build-up'', which is a
convolution of PINUD with an exponential decay function, represented by 
\[
 {\rm PINUD}_{\rm buildup}(t)=\int_{-\infty}^{t}{\rm PINUD}(t_0)\exp{\left(\frac{t_0-t}{\tau}\right)}dt_0
\]
Various values of time constant $\tau$ was tried between 5000--10000 sec
and $\tau=8000$ sec was selected in the latest model.
Typical light curves of PINUD and PINUD$_{\rm buildup}$ are shown in
figure \ref{pinudbuildup}.

Then, PINUD and PINUD$_{\rm buildup}$ are utilized to sort and refer to
the earth occultation data base.
We accumulated PIN event data and pseudo event data (for dead time
correction) under the conditions that the target elevation angle from
the earth limb should
be $<-5^{\circ}$.
Since the PIN NXB count rate was affected by the change of bias voltages
(500V to 400V)
and there is a gradual variation of the NXB, the data base is
accumulated by considering these issues.
There are 16 well unit counters in the HXD, each of which contains 4 PIN
diodes.
Since the NXB variation is somewhat different among units \citep{kokubun07}, the data base
of each unit is sorted and referred to PINUD and PINUD$_{\rm buildup}$ 
of the corresponding unit.
The data base consists of $40\times220$ PIN NXB spectra and pseudo
events for 4 PINs in each unit.
The range of the PINUD in the data base is from 2.5 cts sec$^{-1}$ to 42.5 cts
sec$^{-1}$ per one unit, and that of the PINUD$_{\rm buildup}$ is 
from 4.0$\times10^4$
cts to 2.25$\times10^6$ cts per one unit.
Figure \ref{bgdasort} presents an example of the NXB data base 
divided into four energy bands for the visualization.
It has been found that the long-term variation of the PIN NXB is
significant and it is not so well reproduced by the data base prepared
as above.
Therefore we correct the total PIN NXB count rate in the data base 
by a 2nd order polynomial function of time.

The PIN NXB for each observation is estimated by picking up a spectrum
from the data base, based on the two parameters.
The reference to the data base is performed at each sampling rate of PINUD
(2, 4, 8, or 32 sec).
The exposure of the picked-up spectrum is corrected with dead-time.
A time to the next event is determined by the Poisson statistics, based on
the total count rate of the NXB spectrum.
In order to reduce the statistical error of the
number of events in the PIN NXB model, we apply 10 times as high a PIN
background rate as the prediction.
A pulse height is obtained by generating
random numbers which follow the referred spectrum as a probability
distribution.
These processes are repeated until the event time passes the end time 
of each PINUD sampling period, and the process goes to the event generation 
in the next PINUD sampling period.

This method can generate the NXB model as soon as possible, after 
the PINUD history becomes available.
Therefore, the HXD team supplies the NXB model of this method quickly so
that observers can start the analysis at the same time as the data
become available.
Note that this modeling is not available for the data from 2006 March 23
to 2006 May 13, because of the systematic change of the PIN-NXB count
rate as described in \S2.


\subsection{Method based on Parameterization of background light curve  : {\tt LCFITDT} (PIN/GSO NXB model)}

The GSO NXB exhibits much stronger variations
as a result of activation induced during the SAA passages.
These variations are strongly energy-dependent,
because many activated nuclei emit various gamma-ray lines
together with the continuum.
Since the method of simply using the parameter-sorted data base 
is very difficult to model the
GSO NXB, an alternative method is developed, which directly
fits the NXB light curves with an empirical model, determines its parameters,
and predicts the NXB count rate at any given time.
Note that this modeling ({\tt LCFITDT})
\footnote{This method is referred as ``tuned'' or ``bgd\_d'' 
in the Suzaku SWG team.
The event FITS files generated with this method are identified by the
keyword of {\tt LCFIT} or {\tt LCFITDT} in the {\tt METHOD} record of
the FITS file header.
Both are basically the same method, 
but the former does not include the dead-time correction.}
can be applied to the PIN NXB as well.
In order to reproduce the energy dependence of the background light
curve, we model the background separately in 32 energy bands for the GSO NXB.
On the other hand, the low count rate of the PIN NXB does not allow us
to study the energy dependence of the light curve in detail, and thus we
at first model the light curve in 12--70 keV band and distribute the count
rate in a single 12--70 keV band into 256 energy channels,
based on the data base of PIN NXB spectral shapes
which are sorted by COR\footnote{Here we do not use the PINUD, because
of technical convenience.} and T\_SAA\_HXD (elapsed time after SAA).
However, PIN NXB spectral change would depend on other parameters than the
above two parameters, but it is found that the dependence is too small
to recognize in the early phase due to low event statistics of PIN NXB,
especially for the higher energy band.
Therefore, we included such effects after the first-version modeling
function is constructed and studied (\S\ref{bgddpin}).

Since the modeling needs the complete PINUD history and the
reprocessed gain-calibrated GSO data,
the GSO reprocess becomes available after the release of the 
gain history file for the corresponding observation date, and it is
usually 1.0--1.5 months after the pipeline processing.
Therefore, the NXB model with the {\tt LCFITDT} method is released after
1.5--2.0 months after the pipeline processing, unlike the {\tt
PINUDLCUNIT} method.

\subsubsection{Parameterization of the NXB model}\label{parameterization}

In order to study the time variation of the NXB,
we accumulate the earth occultation data,
under the following conditions.
The target elevation angle should be  $<-5^{\circ}$
(namely, during the Earth occultation);
the data rate of the Satellite Data Processor (DP) should not be low (L);
the in-orbit HXD data transfer should not be saturated;
and the COR should be $>6$ GV.
For the GSO, we divide the accumulated events into 32 energy bands,
and derive 200-sec bin light curves in each energy band.
Boundaries of the 32 energy bands are
logarithmically spaced from 53 keV to 1024 keV,
with each band having a typical BGD rate  of 0.5--1 cts sec$^{-1}$.
The PIN light curve is analyzed in a single 12--70 keV energy band.

After the 32 GSO NXB light curves and one PIN NXB light curve are prepared,
we fit them individually,
with an empirical model to be constructed in the following manner.
Like in the modeling method {\tt PINUDLCUNIT},
the PINUD dependence is represented by
a term which is a second order polynomial function of the
PINUD counts summed over the 64 PIN diodes.
Likewise, the activation component is represented
by PINUD build-up counts.
In this model, several exponential decay functions with different time 
constants $\tau_k$ are considered.
The convolution integral is calculated up to $30 \tau_k$,
in order to save the calculation time.

Various studies indicate
that we need at least 3 or 4 time constants, $\tau_k$,
to represent the build-up effects in each energy band.
In order to find them,
we performed a preliminary analysis.
That is, we fitted a whole light curve of
the Earth occultation data acquired in the period 
from 2005 September 2 to 2006 February 28.
with a model consisting of the second order polynomial function of 
the PINUD counts
and the PINUD build-up term with two time constants.
Then, we produced confidence contours for the two time constants
by scanning them independently.
First, we scanned the two  time constants in a range shorter than one day,
and determined them.
After fixing these two short time constants,
we newly added two build-up  time constants  longer than one day,
and repeated the search.
A shorter time constant was scanned at grid values of
0.5, 0.8, 1.0, 1.5, 2.0, 2.5, 3.0, 4.0, 5.0, 7.0, 10.0,
13.0, 15.0, 18.0, 22.0, 30.0,
45.0, 60.0, 80.0, and 100 ksec,
while a longer one is scanned over 0.15, 0.25,
0.35, 0.45, 0.55, 0.8 1.0, 2.0, 3.0, 5.0,
6.5, 8.0, 10.0, 20.0, and 30.0 Msec.
Thus, we have obtained 4 time constants
that describe the PINUD build-up effects in each energy band.
The obtained four time constants are typically 1--2 ksec,
10--20 ksec, a few days, and several tens days,
depending on the energy band.

While the activation effects have thus been described approximately
by PINUD build-up with four time constants,
the modeling is found to be incomplete, mainly 
due to the following two reasons.
We have found that the activation BGD also depends on the
angle $\theta_{\rm B}$ between the geomagnetic field and the HXD field of view.
When $\theta_{\rm B}$ is small,
the SAA particles directly enter the tight HXD  shield ``Wells''.
As a result, the GSO background count rate due to the activation becomes
higher even for the same entrance number of the SAA particles or
PINUD build-up count
(5--10\% higher when the angle is around 0$^{\circ}$).
When we include this angle dependence, 
the background reproducibility was somewhat improved.
Apart from the $\theta_{\rm B}$ dependence, there might be other
components of activation with time constants other than four time constants
included in the model.
Such components can be represented by real GSO NXB count rate.
Then, we additionally consider the 450--700 keV GSO count rate, 
$GSOHCNT(t)$,
where celestial signals in such energy band are negligible, less than 0.2\% 
even during the Crab observation.

From the above consideration,
the empirical model describing the light curve
in the $i$-th energy band is expressed as
\[
\begin{array}{rl}
  BGD_i(t) =
             & \displaystyle{ a_i 
             + \sum_{k=0}^3 b_{k,i}\int PINUD(t') \cdot 
\exp{\left(-\frac{t-t'}{\tau_{k,i}} \right)}dt'} \\
             & \displaystyle{
             + \sum_{k=0}^3 c_{k,i}
                \left[\int\frac{90^{\circ}-\theta _{\rm B}(t')}{90^{\circ}}  \cdot PINUD(t') \cdot 
\exp{\left(-\frac{t-t'}{\tau_{k,i}} \right)}dt'\right]^p} \\
           & \displaystyle{+ d_i \cdot PINUD(t)+  e_i \cdot PINUD^2(t)} \\
           & \displaystyle{+ f_i \cdot ( GSOHCNT(t) -  GLC(t) ) \cdot \left[ 1+g_i \exp 
\left(-\frac{t-t_{\rm SAA}}{\tau_g} \right) \right]} \\
           & + h_i(t)~~,
\end{array}
\]
where the coefficients $a_i$, $b_{k,i}$, $c_{k,i}$, $d_i$, $e_i$, $f_i$, and $g_i$
are model parameters to be adjusted,
and $t_{\rm SAA}$ is the elapsed time from the end of the latest SAA,
while $\tau_{k,i}$ are fixed to the values as obtained above and $\tau_g$
are fixed to 10000 sec as described later.
The index $p$ is 1, except that it is 2 for two shorter time scales 
of the PIN NXB.
The terms with the coefficients, $b_{k,i}$ and $c_{k,i}$, represent
the activation, and the term with $d_{i}$ and $e_{i}$ does the PINUD
dependence.
The term with $f_{k,i}$ and $g_{i}$ is included to reproduce the
activation more accurately with $GSOHCNT(t)$.
$GLC(t)$ is a gradually increasing function of the GSO count rate, and obtained
by fitting the light curve of 450--700 keV band from 2005 Aug 17 to the
current time by the PINUD-build up terms with two long time constants of 0.35
Msec, and 30 Msec.
This function subtracts the long-term gradual increase from $GSOHCNT(t)$
so that the GSO count rate only represents the
day-by-day background variation that cannot be fully reproduced by other model
components.
Furthermore, to consider the difference of the time constant of
activation between 450--700 keV and other energy band, we include the
exponential term with a time constant of 
$\tau_g =10000$ sec, which is found to give a good reproducibility.
The last term in this model, $h_i$, is  a correction bias
to be explained later,
introduced in order to reduce the
current uncertainty as much as possible.
As a result, input data for modeling the background
are $PINUD(t)$, $GSOHCNT(t)$, $\theta_{\rm B}(t)$, and $t$.

\subsubsection{Modeling Procedure of the GSO NXB}\label{bgddgso}

Here we describe the procedure of the GSO NXB modeling, based on the
model function in \S\ref{parameterization}.
The procedure of the PIN NXB modeling is somewhat different, and thus we
describe it in \S\ref{bgddpin}, especially for the specific issues.

A set of monthly model parameters are determined
by fitting the light curve of the Earth occultation data from each month,
together with those before and after 10 days of that month.
Time region of each fit does not cross any occasion
when the HXD operation mode (such as high-voltages
and lower discriminator levels) was changed.
After once performing the fit,
we exclude data points with large deviations
by $>5\sigma$ ($\sigma$ is a root-mean-square),
and perform the fit again to obtain the final parameters.
In the fitting, we fix the correction bias $h_i$ to 0.
Afterwards,
we calculate the residual between the background and the model in 
every 150 ksec,
and employ the residual as the correction bias  $h_i$;
$h_i(t)$ varies every 150 ksec,
while the other model parameters are constant in each month.
Here we adapt 150 ksec, since residual in a shorter time scale picks up 
the Poisson
fluctuation and one in a longer time scale smears out the residual profile.
Typically, $h_i$ is at most 0--2\% of the total background.
Currently, the number of parameters is still not optimized, and therefore
parameter couplings in the light curve fitting sometimes introduce 
a small discontinuity of count rate in each component at the month
boundary, but the discontinuity is canceled after adding each component.

\begin{figure}[hpbt]
\centerline{\includegraphics[width=8cm]{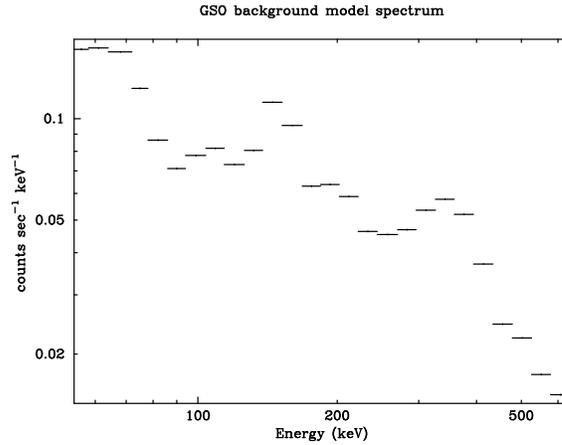}}
\caption{An example of  the GSO background model spectrum.}
\label{bgdgsospec}
\end{figure}

After the BGD model parameter sets
are thus determined for each energy band,
we create the background light curve in each energy band at each PINUD sampling
time.
In this process, we correct the dead-time by using the dead-time light
curve estimated from the HXD pseudo event
\footnote{Currently, the GSO NXB model is not corrected by the dead-time,
and thus we refer it as the {\tt LCFIT} model, to distinguish the
dead-time corrected PIN NXB model.
In the near future, we will release the dead-time corrected {\tt
LCFITDT} GSO NXB model.}.
Finally, BGD events in each band are created,
with their pulse heights determined by a Monte-Carlo method
referring to the model-predicted counts in each PINUD sampling period.
As shown in figure \ref{bgdgsospec},
the pulse height is uniformly and randomly distributed
within each energy band of the GSO.
Therefore, users should use
exactly the same energy boundaries as the present model,
when binning the GSO spectra.

\subsubsection{Procedure of the PIN NXB modeling}\label{bgddpin}

The procedure to prepare the model of the PIN NXB light curve is the
same as the GSO NXB, but for only one energy band.
The correction factor $h_i(t)$ is also created in the same way as the GSO
NXB model.
In order to reproduce the PIN NXB spectrum,
pulse height of each event is generated by random numbers, which
follows probability distributions
referring to the actual pulse-height spectral database,
accumulated under various values in the COR and T\_SAA\_HXD;
the database is sorted at boundaries
of 6, 7, 8, 9, 10, 11, 12 GV  in the COR,
and 2000, 4000, 10000 sec with respect to T\_SAA\_HXD.
The database is created from the earth occultation data
during 100 days centering on each month, together with the condition that
the observational
set-up of the HXD is the same within each period.

\begin{figure}[htb]
\centering
\includegraphics[angle=0,width=0.4\textwidth]{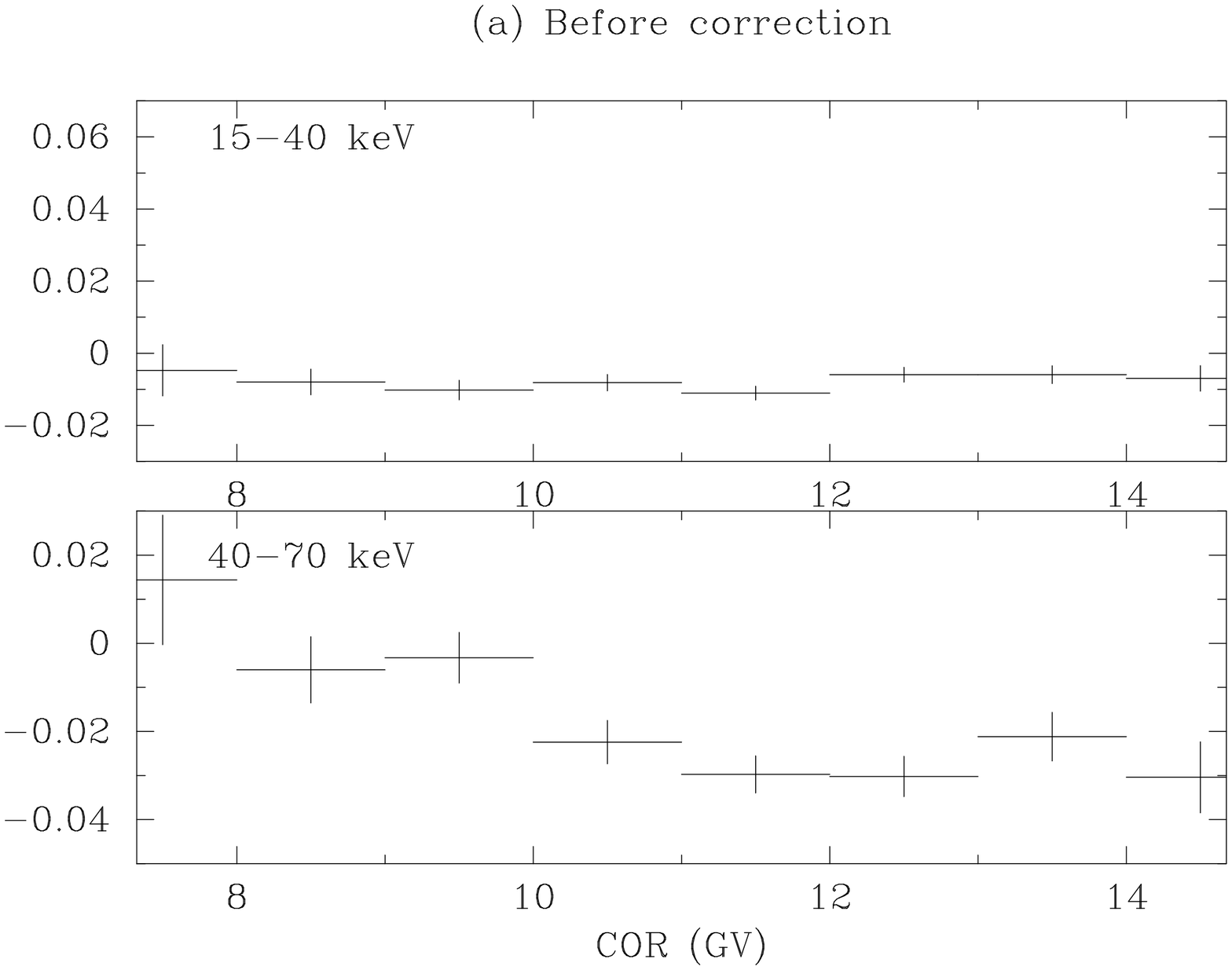}
\includegraphics[angle=0,width=0.4\textwidth]{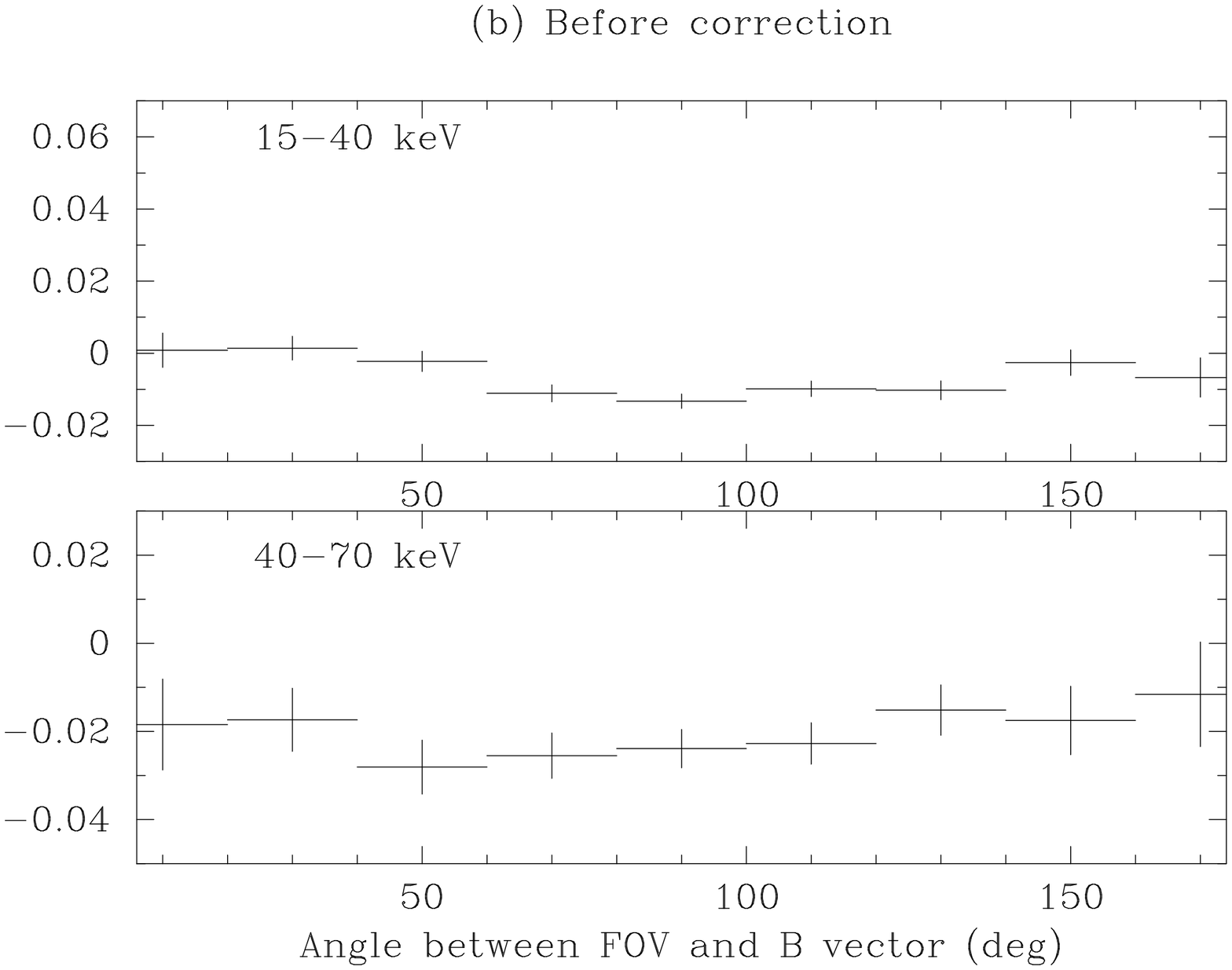}
\includegraphics[angle=0,width=0.4\textwidth]{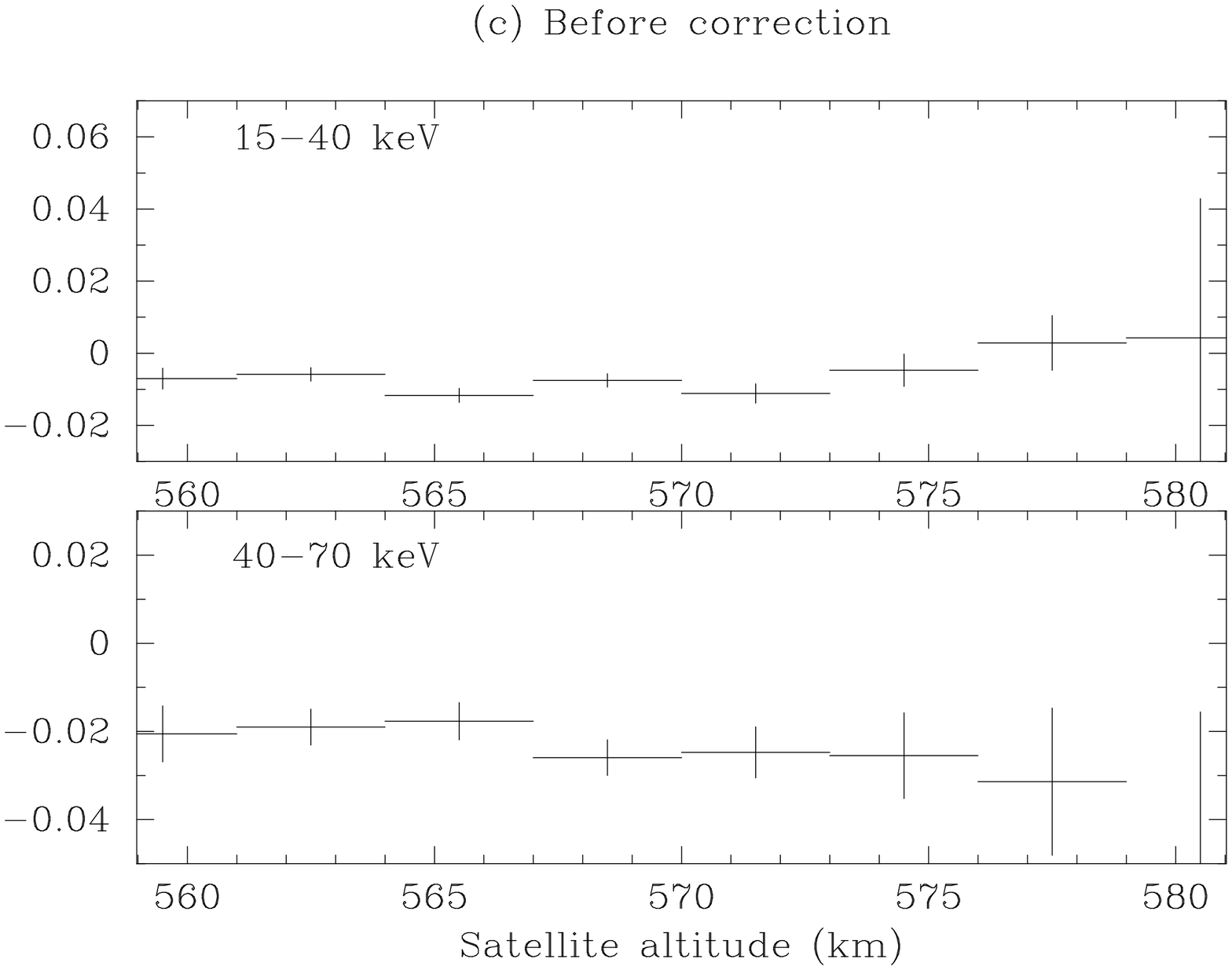}
\\
\includegraphics[angle=0,width=0.4\textwidth]{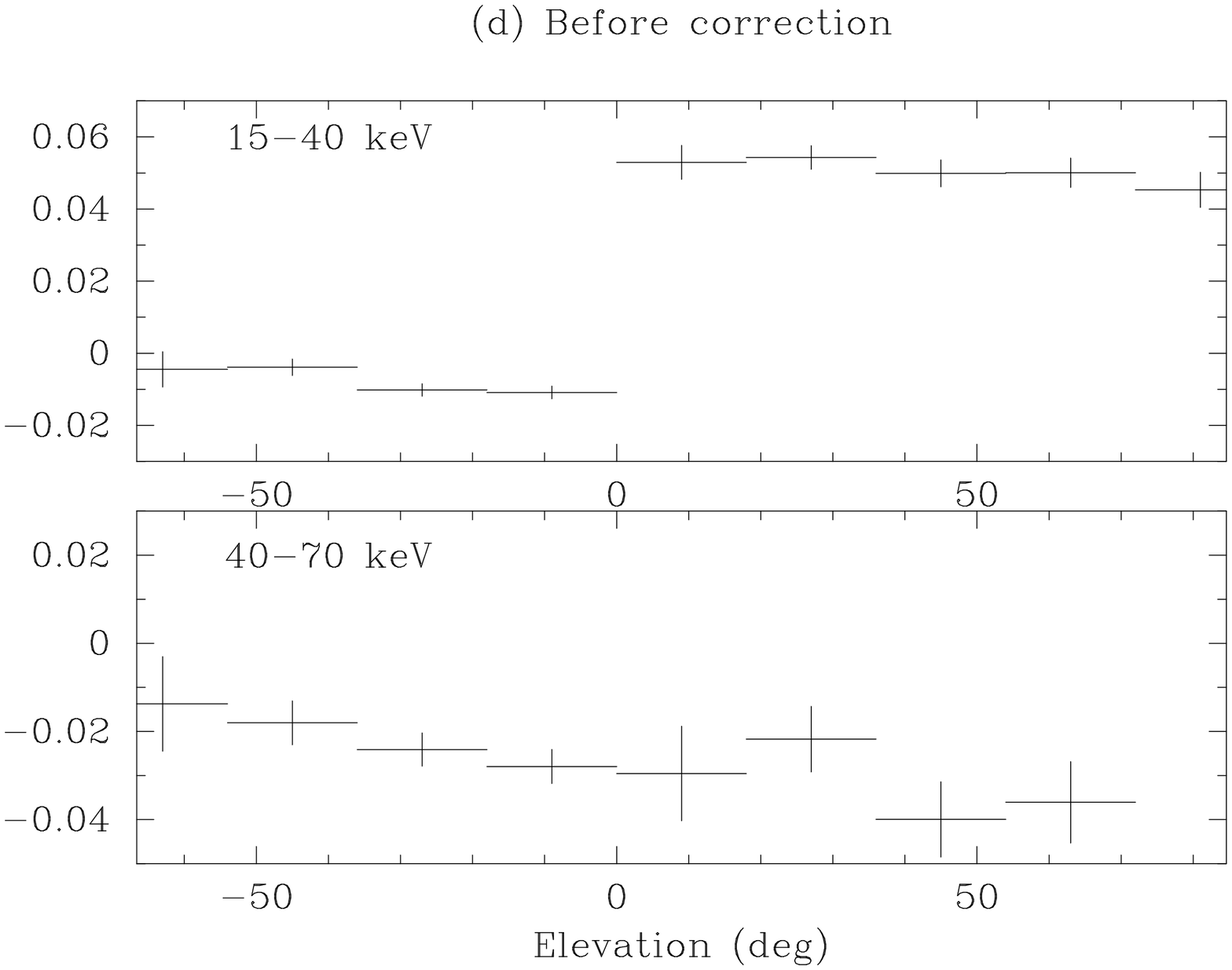}
\includegraphics[angle=0,width=0.4\textwidth]{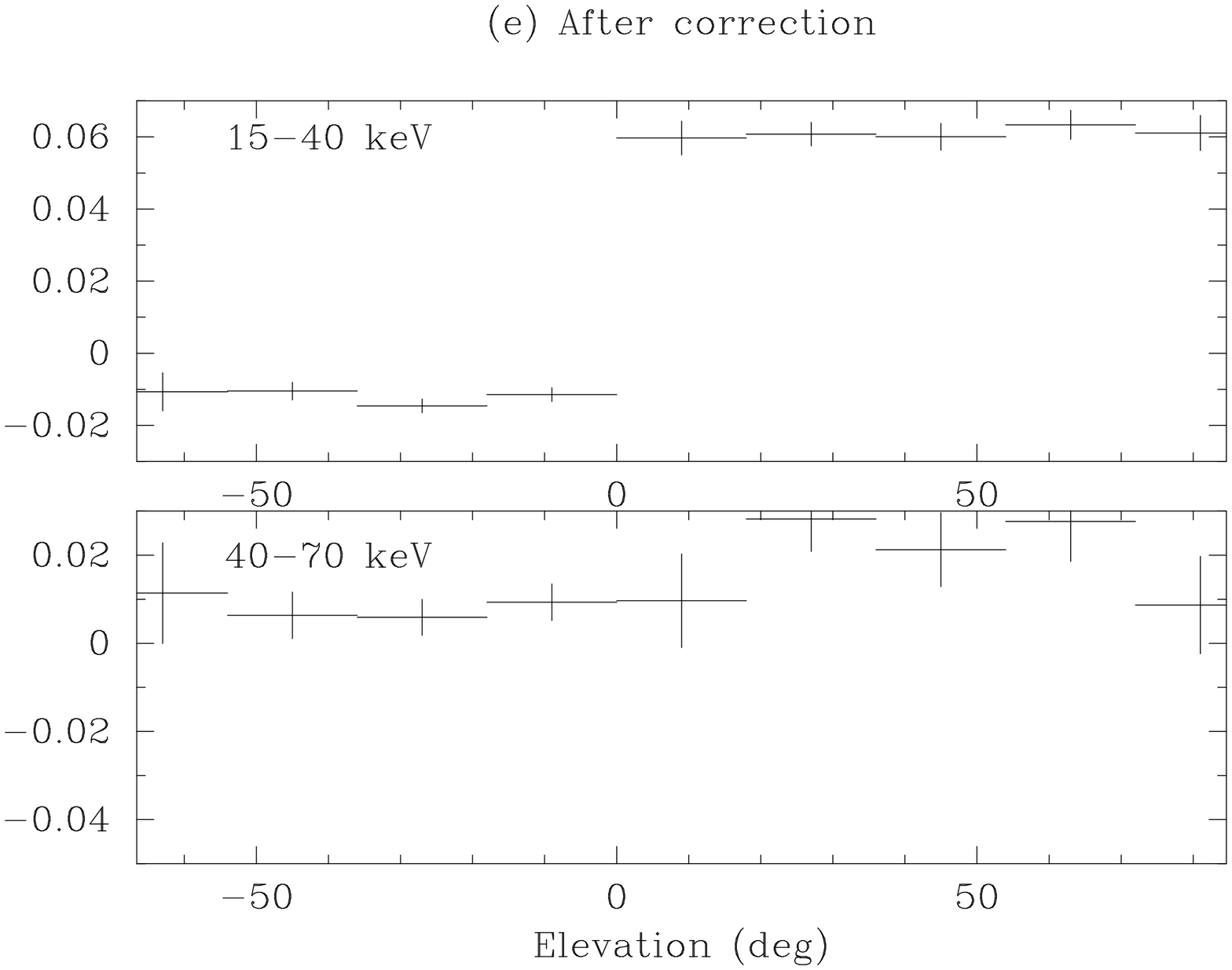}
\caption{
The residual of PIN background subtraction against the (a) COR, (b) $\theta _B$,
 (c) satellite altitude, and (e) earth elevation angle before the
 correction of background model.
Exceptionally, 
the bottom-right figure (e) shows the residuals against the earth elevation
after the correction of background model. 
}
\label{elev1}
\end{figure}

Although we represent the energy-dependence of the PIN NXB light curve
as described above, the modeling accuracy is limited because we consider
only two parameters to reproduce the energy dependence.
The PIN NXB count rate is dominated by events in lower energy band, and
therefore the above modeling is relatively good for lower energy band,
but the accuracy becomes worse toward the higher energy band.
If the model shows systematic deviations from the
real NXB data and the deviations can be modeled with some
appropriate parameters, we can improve the
reproducibility of the PIN NXB model.
Figure \ref{elev1} and \ref{elev2} 
show an example of comparison of the NXB model with the earth data
against the elevation angle from the earth rim, the COR, the satellite
altitude, and the angle $\theta_{\rm B}$ between the geomagnetic field
and the Suzaku field of view.
Here we averaged the earth occultation data from 2005 September to 2007 August.
In the plot against the elevation angle, 
the blank sky data defined in \S4 are used to cover the elevation
above $0^{\circ}$.
Systematic residuals of several \% are clearly seen in these figures,
but they can be recognized for the first time after long
accumulation of the PIN NXB data,
bacause of low count rate of the PIN NXB, especially 
in a higher energy band.
The residuals against the COR are due to the incomplete modeling of the
COR dependence by PINUD.
Toward the higher satellite altitude, the residuals change positively
or negatively in lower or higher energy band of the PIN NXB, respectively,
probably because the particle environment becomes different along the altitude.
Negative correlation is seen in the residual against
the elevation in 40-70 keV band; the model tends to
overestimate the NXB at larger elevation.
Note that the jump at the elevation of $0^{\circ}$ is due to the CXB
signal above $0^{\circ}$.
Since the earth albedo is at most 0.3 \% of the PIN NXB, 
the elevation dependence cannot be explained by it.
Perhaps this dependence is thought to be due to secondary 
particles created through the
interaction between primary cosmic rays and earth atmosphere, and
intensity of secondary particles is somewhat higher from the earth side.
The residual in 15--40 keV becomes larger for the $\theta_{\rm B}$ 
around $30^{\circ}$
and $150^{\circ}$, possibly because the effective thickness of the
BGO active shield becomes the largest and the NXB becomes smaller for
these incident directions.

Then, we corrected the above systematic dependences against the
elevation, satellite altitude, COR, and $\theta_{\rm B}$.
Residual profiles are obtained in the energy range 
of 15--25, 25--40, 40--70 keV and are fitted with a 3--5th other
polynomial
function of the corresponding parameter (COR, elevation ...).
The correction factor for arbitrary energy is calculated by linear
interpolation between center energies of 3 energy band.
Even after that, the comparison of the PIN NXB spectrum between the
earth data and the model has an energy-dependent systematic residual at
$\sim$2 \%, and thus we corrected it furthermore by a 4th other 
polynomial as a 
function of energy (the term $K_5(j)$ in the following formula).
As a result, we obtain the PIN NXB light curve as,
\[
 BGD_j^{\prime}(t) = BGD_{12-70 {\rm keV}}(t)K_1(COR,j)K_2(H_{\rm sa},j)
K_3(\theta_{\rm elev},j)K_4(\theta_{B},j)K_5(j)F(j)
\]
where $j$ is the pulse height channel of the PIN event,
$BGD_j^{\prime}(t)$ is the model count
rate in the $j$-th pulse height channel, $BGD_{12-70 {\rm keV}}(t)$ is
the model count rate in 12--70 keV band corresponding to $BGD_i(t)$ 
in \S\ref{parameterization},
$H_{\rm sa}$ is
a satellite altitude, $\theta_{\rm elev}$ is an earth
elevation, $K_{1,2,3,4}$ is a
correction function, and $F(j)$ is a probability function of pulse
height.
In figure \ref{elev1}, the residual against the elevation angle for the
corrected PIN NXB model is shown, demonstrating that 
the elevation dependence becomes reduced.

\clearpage
\section{Reproducibility of the PIN NXB}

\subsection{Data Reduction}

In this section,
we examine the reproducibility of the NXB by utilizing 
the available sky and earth occultation data in 2005 Aug 17 to 2008 Jan 31
for the ver 2.0 pipeline processing.
The event selection criteria are the same
as those of cleaned event. To be specific, we applied the
following selection criteria
\footnote{In the Suzaku analysis software, the criteria are given as {\tt COR>6} and ({\tt T\_SAA\_HXD>500 and TN\_SAA\_HXD>180})
and ({\tt HXD\_HV\_W[0123]\_CAL>700 and HXD\_HV\_T[0123]\_CAL>700}) and 
({\tt AOCU\_HK\_CNT3\_NML\_P==1 and ANG\_DIST<1.5}) and ({\tt ELV>5} or {\tt ELV<-5})
}
: the COR is greater than
6~GV, the elapsed time after the passage of
SAA (South Atlantic Anomaly) is more than 500~s and the time
to the next entry to SAA passage is more than 180~s, high voltages
from all eight HV units are in normal value, 
and the satellite is in pointing mode and the satellite attitude is 
stabilized.
Elevation angle from the Earth rim
is more than 5$^{\circ}$ or less than -5$^{\circ}$
for sky and Earth observation, respectively.
We also use {\tt hxdgtigen} 
to discard the time interval when the telemetry was saturated.
Furthermore, the data and background for sky and earth observations 
are accumulated within the same good time interval (GTI).

\subsection{Comparison with the earth occultation data}

\begin{figure}[htb]
\centering
\includegraphics[width=0.49\textwidth]{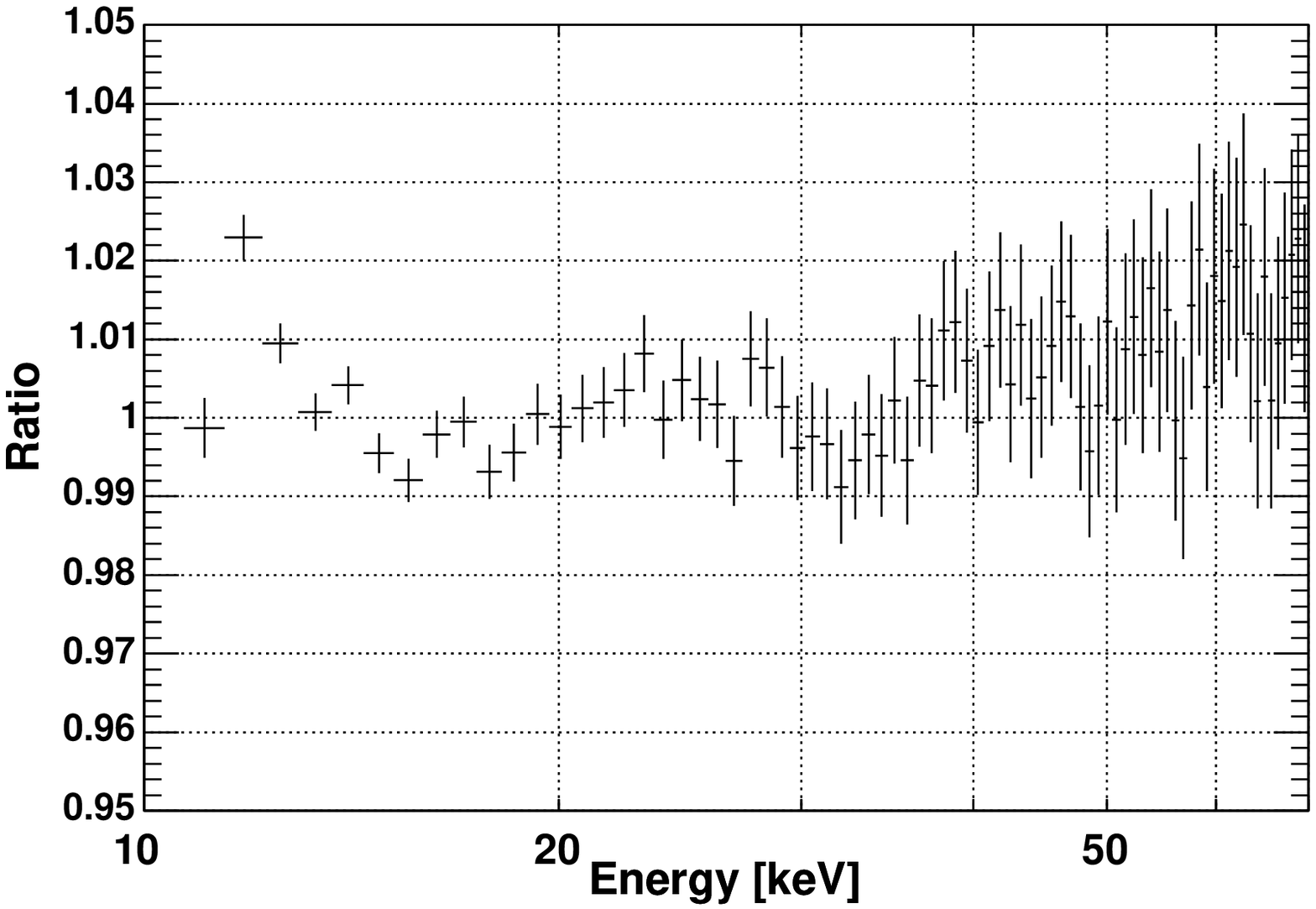}
\includegraphics[width=0.49\textwidth]{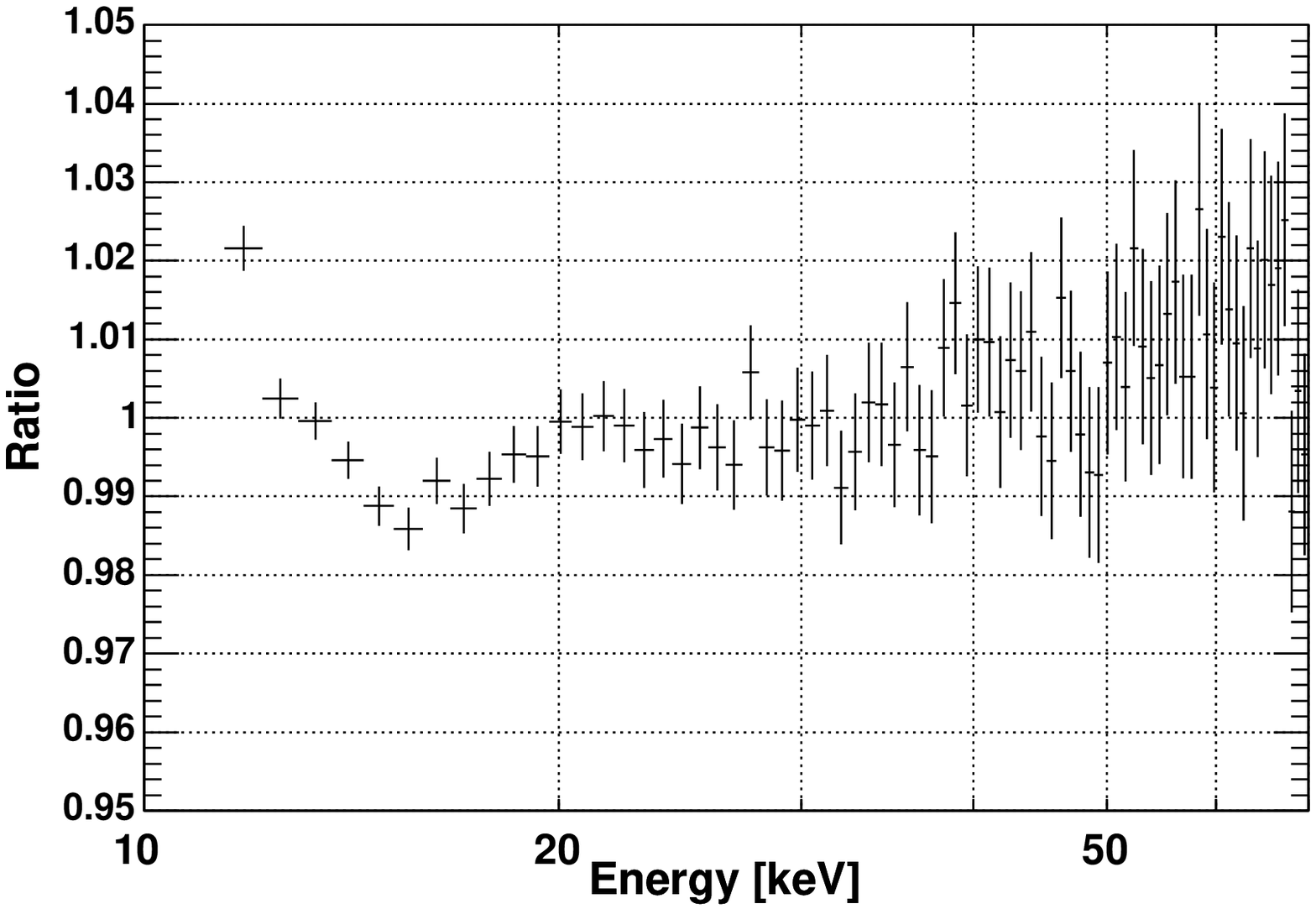}
\caption{
Ratio of spectra between the data and the background model
prediction of all the available Earth occultation data
(7.9~Ms exposure). 
Left and right panels show the plots for {\tt PINUDLCUNIT} and {\tt
 LCFITDT}, respectively.
}
\label{fig:EarthSpectrum}
\end{figure}

Here we show the reproducibility of both two PIN NXB models, by
comparing the model with the earth occultation data.
Although the background model is produced based on the earth data, the
modeling is not complete, and therefore it is important to check how the
modeling itself is accurate, before comparing the model with the sky data.
Figure~\ref{fig:EarthSpectrum} 
compares energy spectra between the data and the NXB model
during the Earth occultation of all available data 
(7.9~Ms exposure in total).
We see a very good agreement
between the model and the data, and can conclude that they agree
within $\sim 1\%$ in 15--70~keV 
for the long-duration averaged Earth occultation data.
Small but systematic deviation is seen below 20~keV.

\begin{figure}[htb]
\centering
\includegraphics[width=0.49\textwidth]{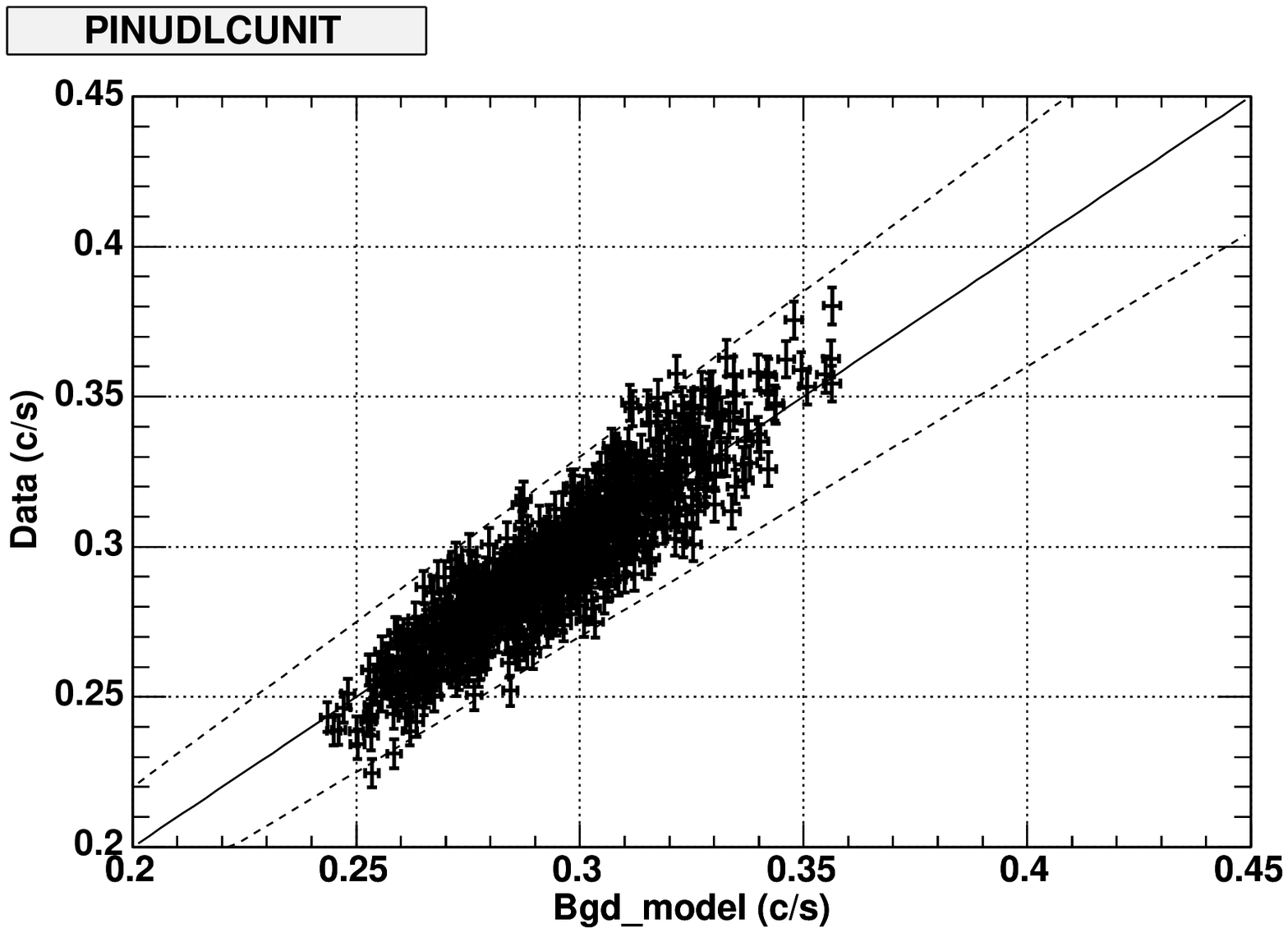}
\includegraphics[width=0.49\textwidth]{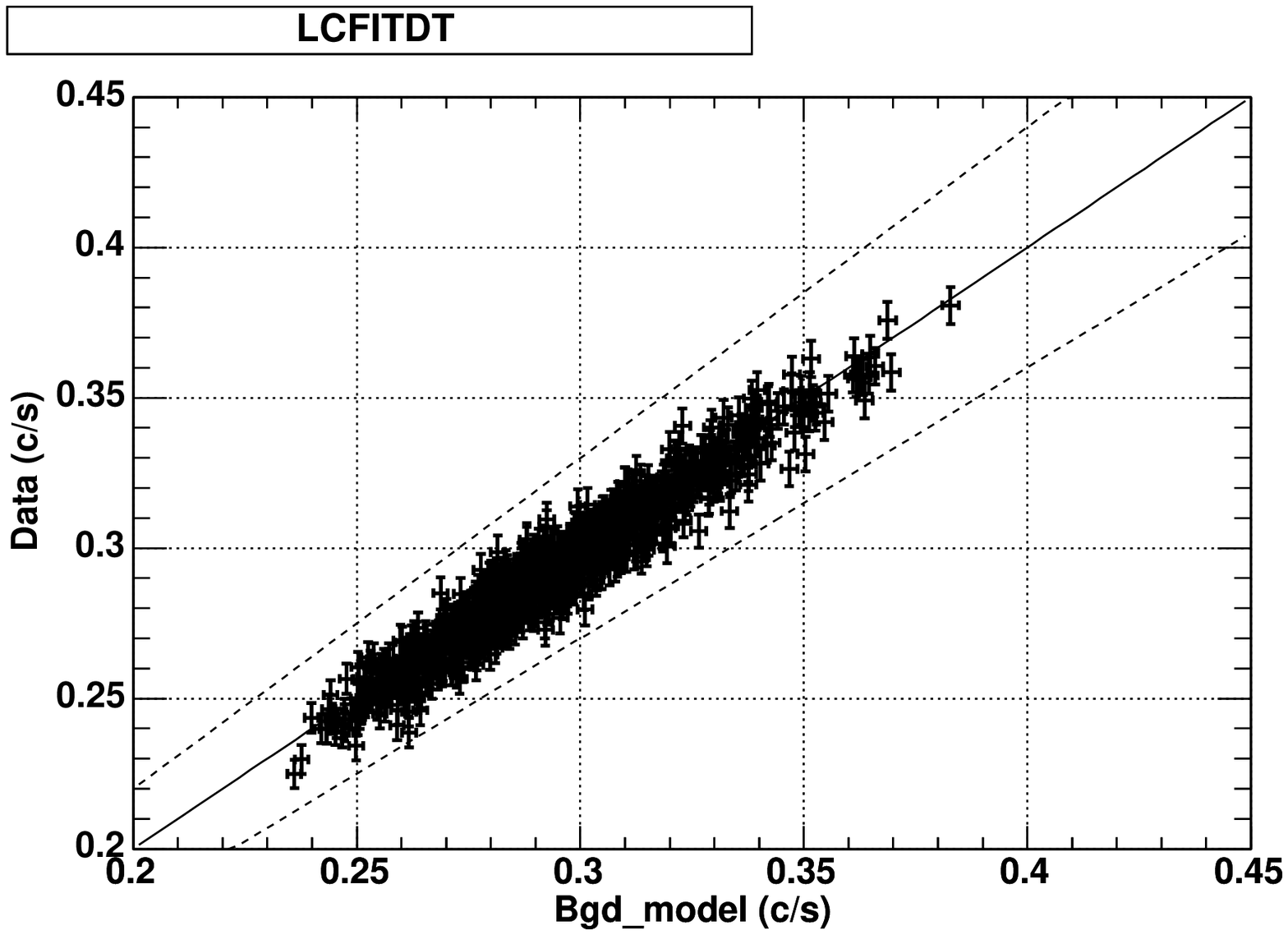}
\includegraphics[width=0.49\textwidth]{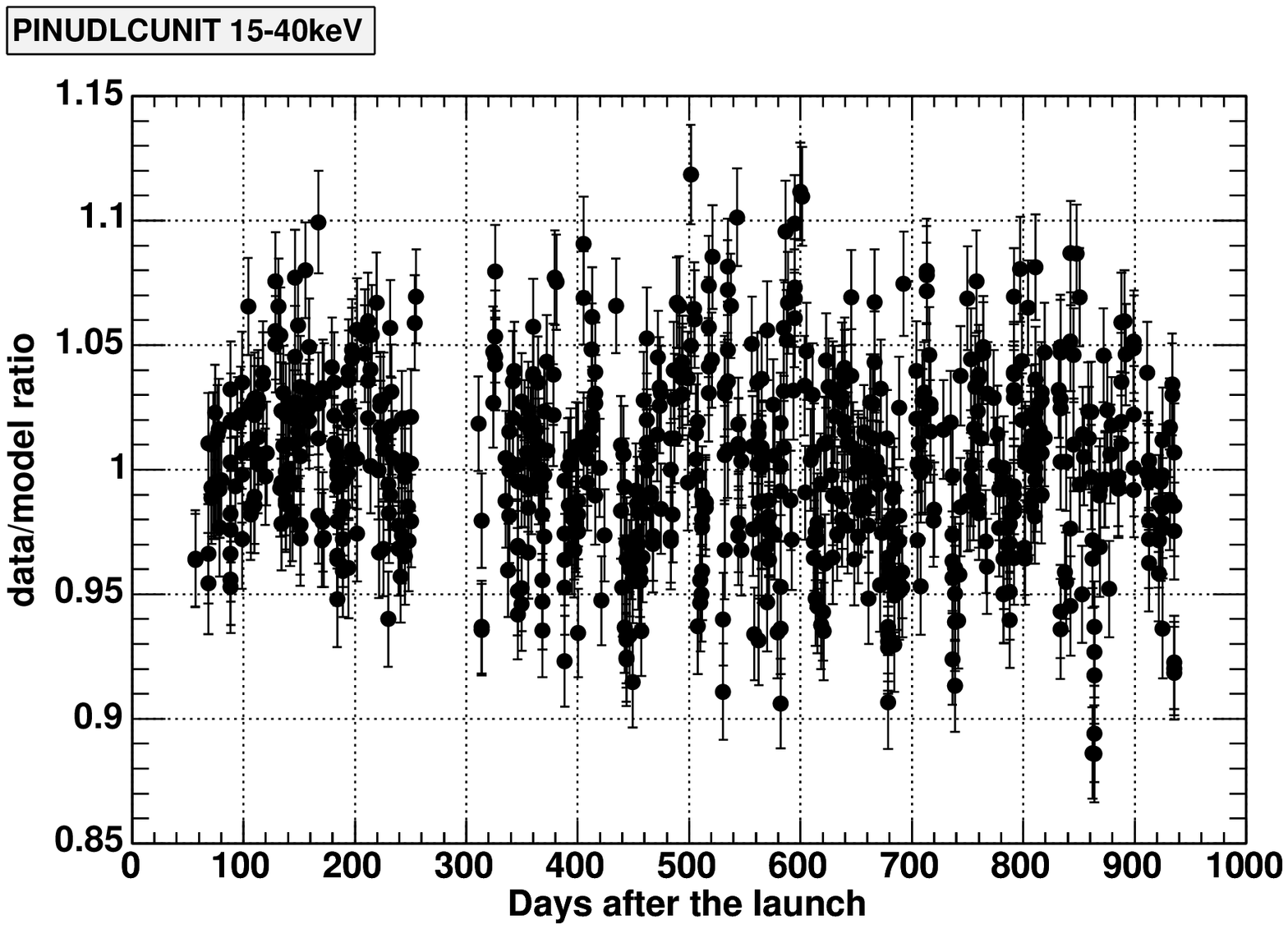}
\includegraphics[width=0.49\textwidth]{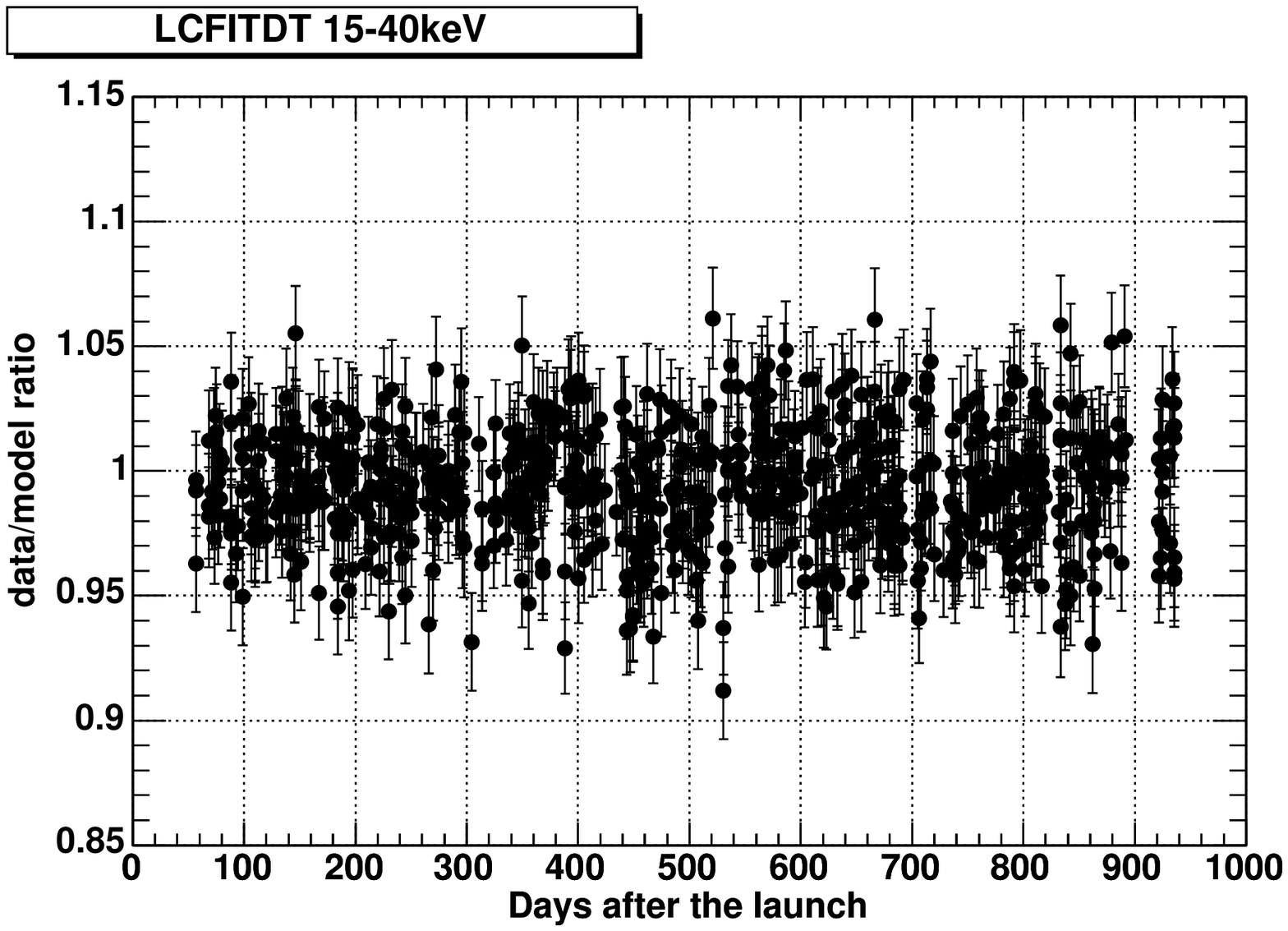}

\caption{Comparison of the NXB count rate in 15--40~keV
between the data and the model prediction. 
Left and right panels show the plots for {\tt PINUDLCUNIT} and {\tt
 LCFITDT}, respectively.
Top panels show a scatter plot between the NXB data and the background
 model. Bottom panels show data to model ratios of the NXB count rate, 
plotted against the day since 2005 July 10 (the day of Suzaku launch).
}
\label{fig:IsobePlotEarth}
\end{figure}

\begin{figure}[htb]
\centering
\includegraphics[width=0.49\textwidth]{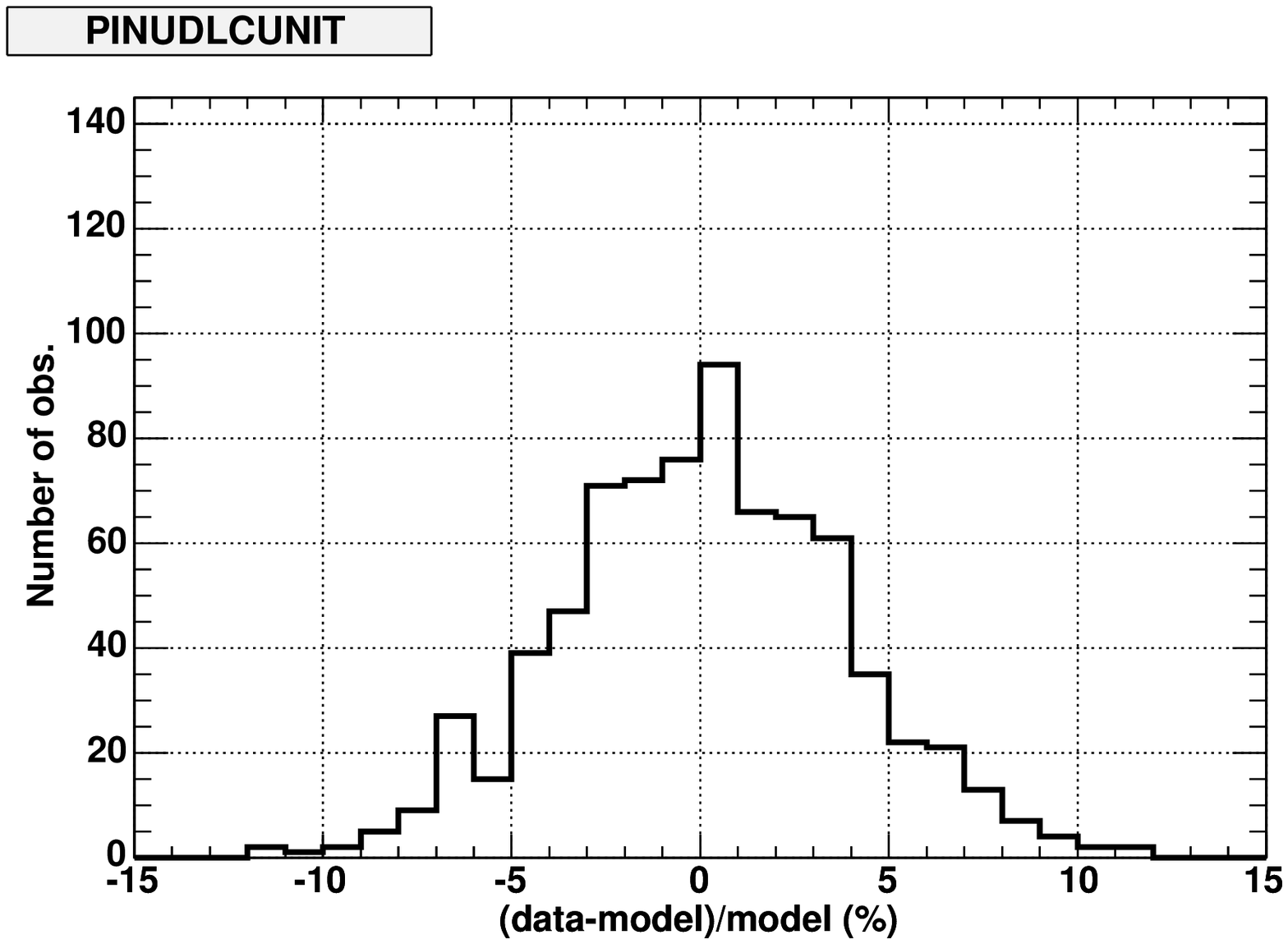}
\includegraphics[width=0.49\textwidth]{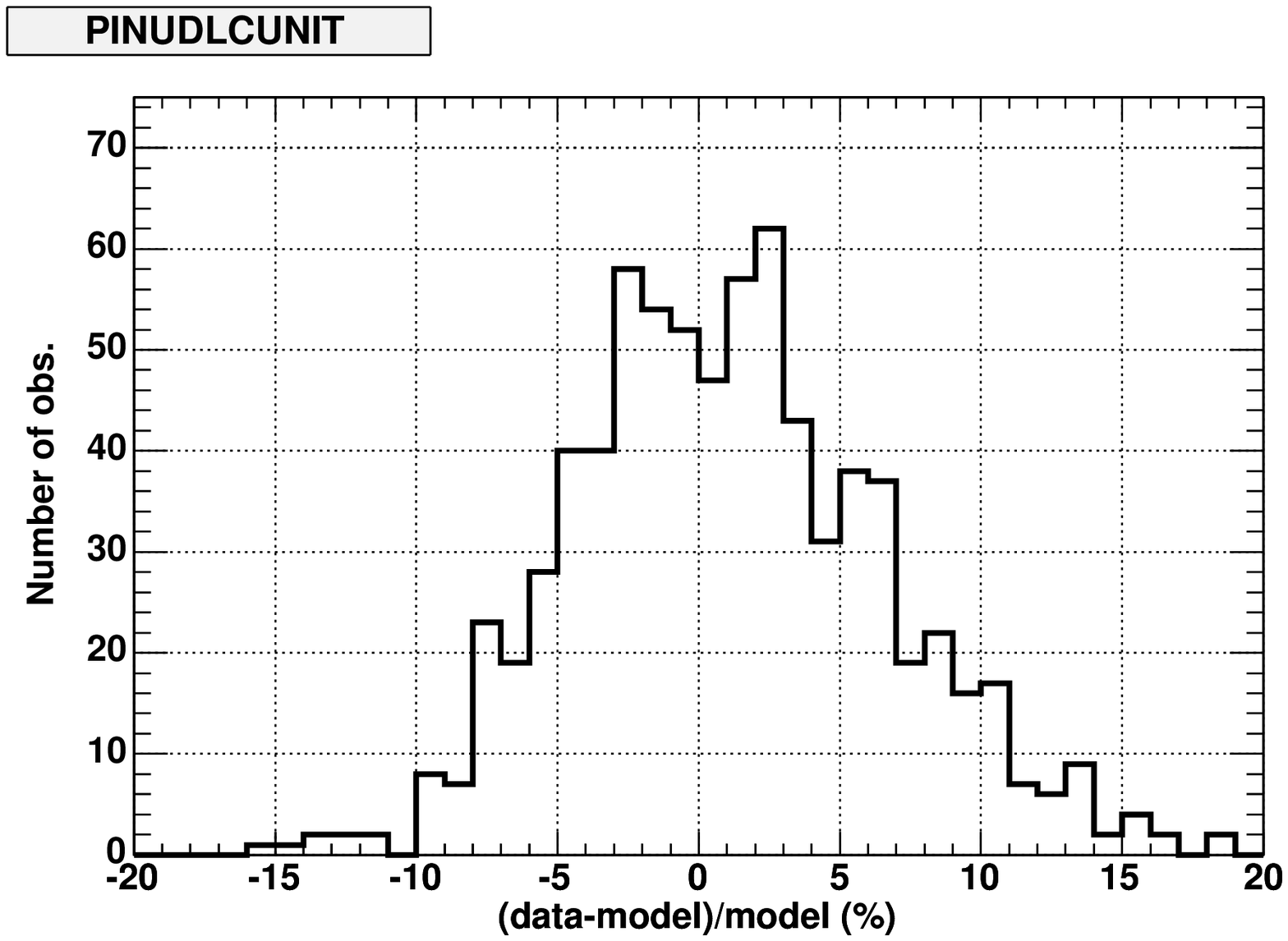}
\vspace*{5mm}
\includegraphics[width=0.49\textwidth]{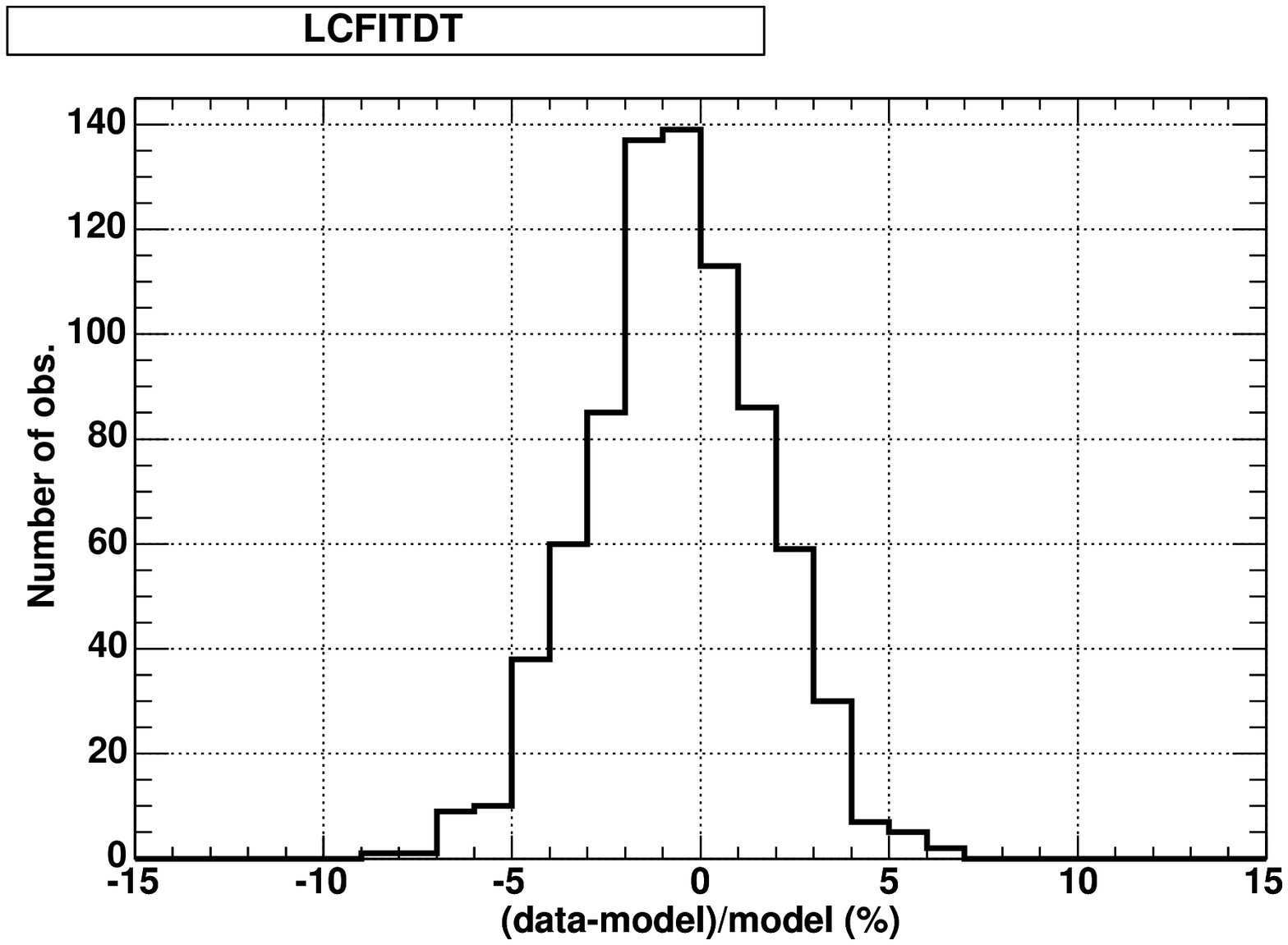}
\includegraphics[width=0.49\textwidth]{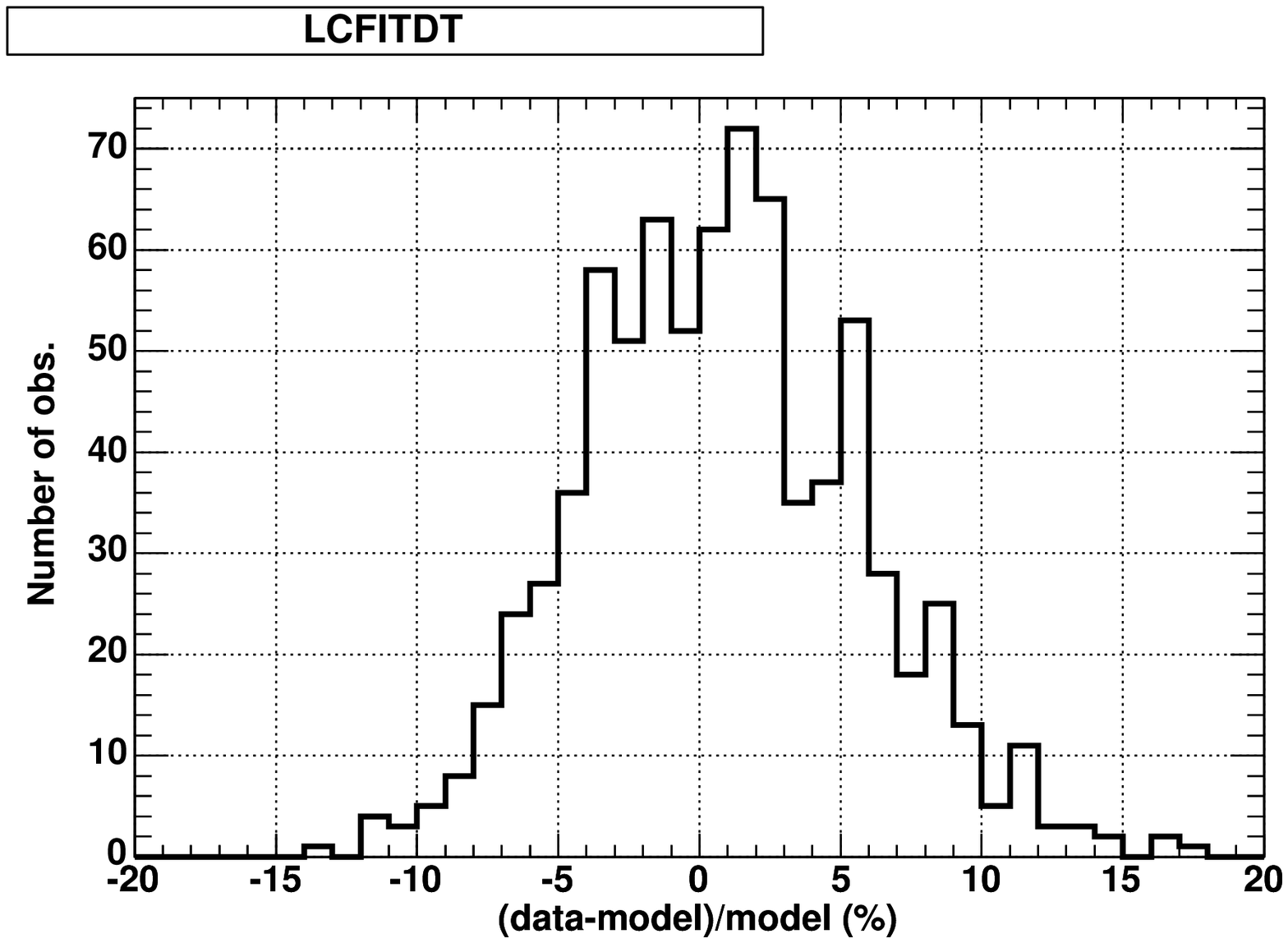}
\caption{Histograms of the fractional residual
 between the NXB data and background model.
Top and bottom panels show the plots for {\tt PINUDLCUNIT} and {\tt
 LCFITDT}, respectively.
Left and right show the plots for 15--40 keV and 40--70 keV, respectively.
}
\label{fig:IsobePlotEarth2}
\end{figure}

For studying reproducibility in a shorter time scale, 
we split the Earth data in each target observation 
into small pieces with 10~ks exposure 
and compared the NXB count rate between the data and the model
in 15--40~keV and 40--70~keV range as shown 
in figure \ref{fig:IsobePlotEarth} and \ref{fig:IsobePlotEarth2},
respectively, and the results are summarized in table \ref{pinres}.
We see that the model reproduces the data in 15--40~keV within
$\pm 7\%$. 
In figure \ref{fig:IsobePlotEarth}, it is found that the reproducibility
is somewhat worse for the {\tt PINUDLCUNIT} in such a way that the model
count rate is overestimated at higher background rate and the data
to model ratio has a large scatter and gradually varies with time.
More quantitatively, the average of the residual is -0.5\% and -0.62\%
and the standard deviation is 3.75\% and 2.31\%, whereas the statistical error
($1 \sigma$) is 1.83\%, resulting in the systematic uncertainty of
3.27\% and 1.40\%, for {\tt PINUDLCUNIT} and {\tt LCFITDT}, respectively. 
Therefore, the systematic uncertainty of the {\tt LCFITDT} ``tuned'' model is 
less than half of that by the {\tt PINUDLCUNIT} ``quick'' model,
and the statistical error now dominates the residual for the {\tt
LCFITDT} model.
We understand the negative mean of the residual
for both models is due to the "dip" around 15~keV in figure~\ref{fig:EarthSpectrum}.
On the other hand, the statistical error dominates the residual in 40--70~keV:
the average of the residual is 1.13\% and 0.81\% and the standard deviation
is 5.53\% and 4.92\% 
for {\tt PINUDLCUNIT} and {\tt LCFITDT}, respectively,
where the statistical error is 4.03\%.
Therefore, the standard deviation of residuals in high energy band
is dominated by statistical errors.

\begin{figure}[htb]
\centering
\includegraphics[width=0.49\textwidth]{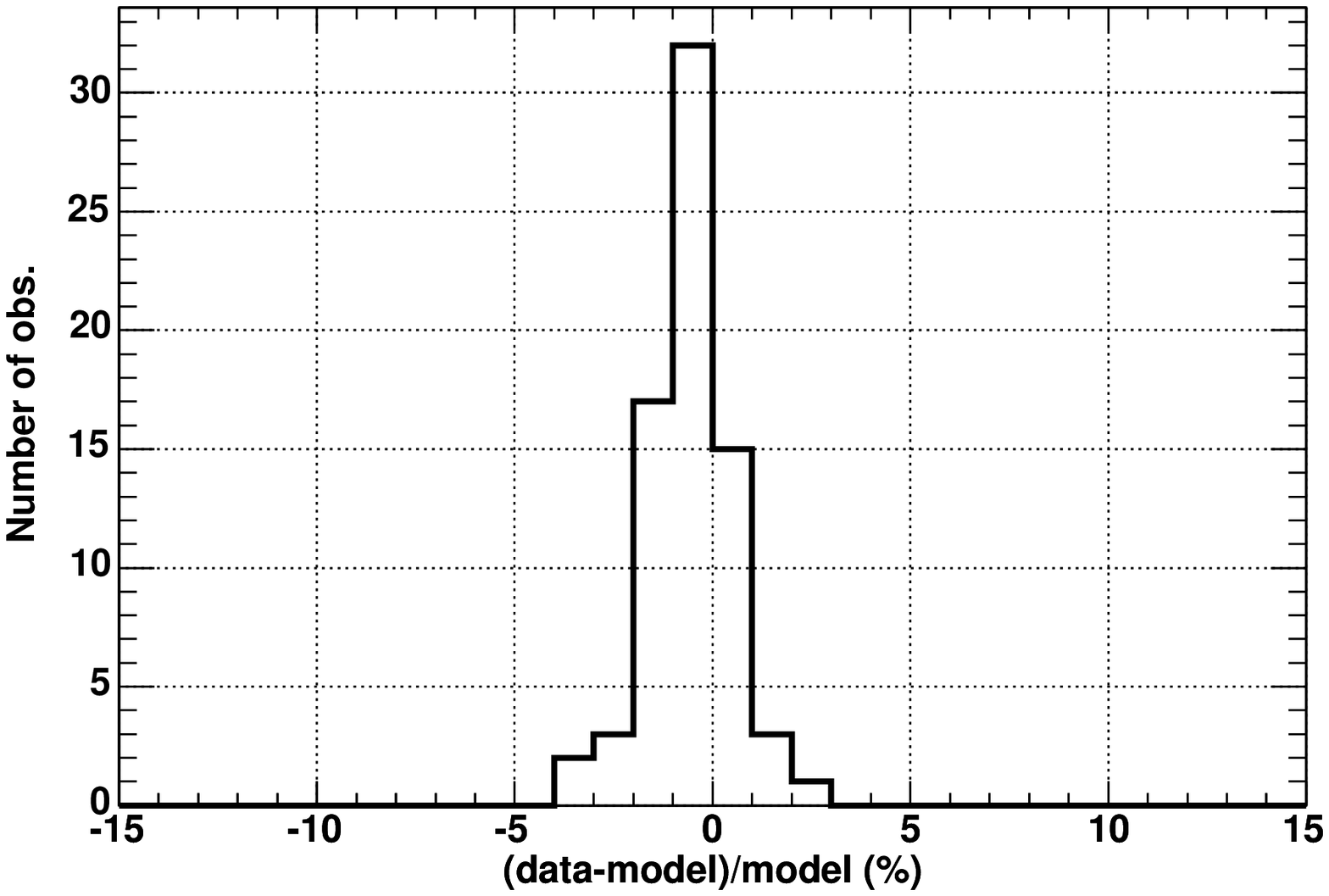}
\includegraphics[width=0.49\textwidth]{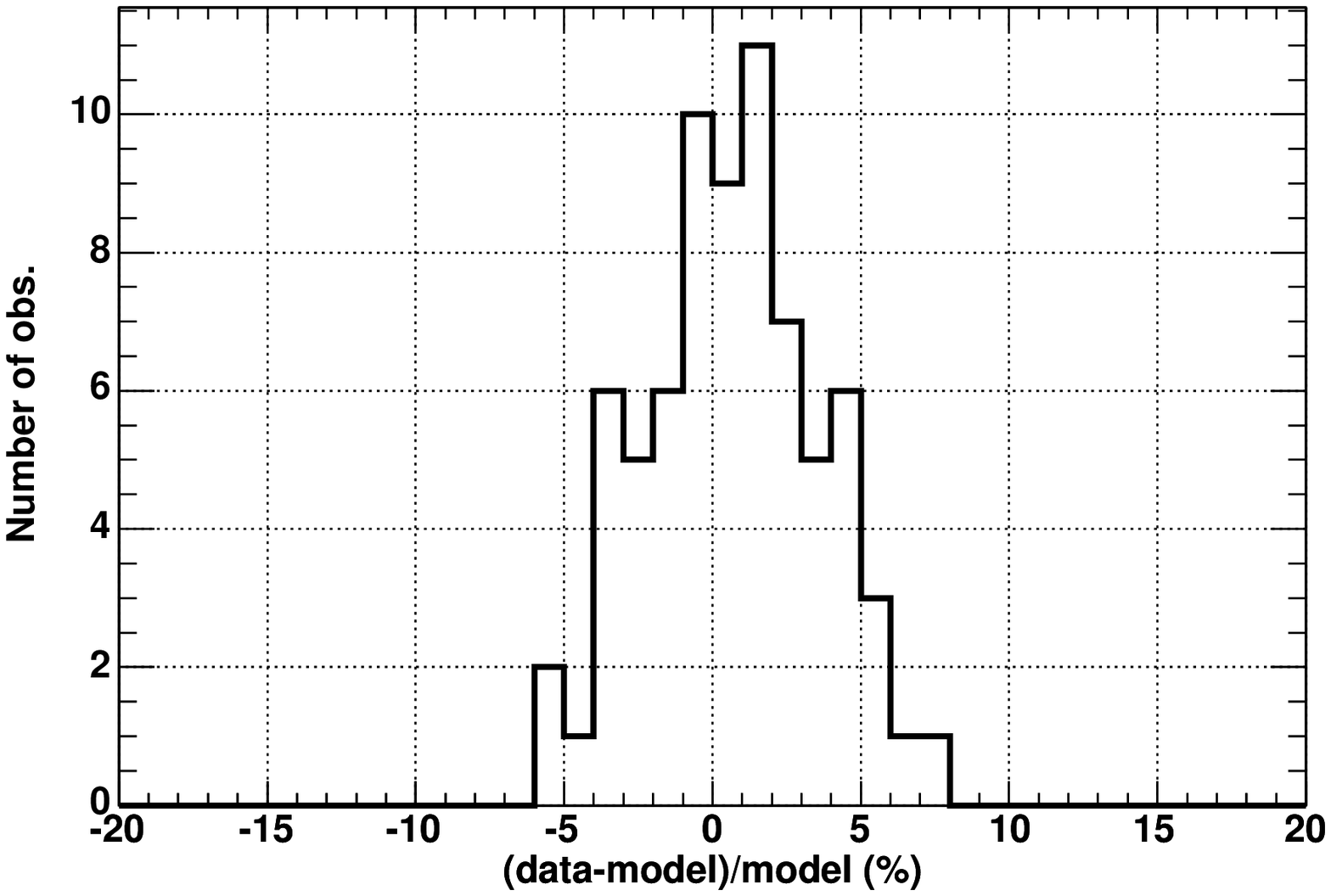}
\caption{The same plots for the {\tt LCFITDT} at those of Figure~\protect\ref{fig:IsobePlotEarth2},
but the integration time is 40~ks instead of 10~ks.
Left and right show the plots for 15--40 keV and 40--70 keV, respectively.
}
\label{fig:IsobePlotEarth3}
\end{figure}

In order to investigate the reproducibility for the {\tt LCFITDT} method
by avoiding the statistical error,
we did the same comparison but with a longer integration time of 40 ks
as summarized in table \ref{pinres} and figure~\ref{fig:IsobePlotEarth3}.
As a result, 
the average, the standard deviation, and the statistical error of residuals
in 15--40~keV is -0.58\%, 0.99\%, and 0.93\%, respectively.
This results in the systematic uncertainty of about 0.34\%. 
For 40--70~keV,
the average of the residual is 0.70\% and the standard deviation and the statistical 
error is 2.87\% and 2.03\%, respectively,
resulting in the systematic uncertainty of about 2.02\%
which is comparable to the statistical error.

\begin{table}[ht]
\caption{Standard deviation (1$\sigma$) 
of residuals between the PIN earth data and model.}
\label{pinres}
\begin{center}
\small
\begin{tabular}{ccccccc}
\hline\hline
15--40 keV & \multicolumn{2}{c}{10 ks}& \multicolumn{2}{c}{20 ks} & \multicolumn{2}{c}{40 ks} \\
 & {\tt PINUDLCUNIT} & {\tt LCFITDT} & {\tt PINUDLCUNIT} & {\tt LCFITDT} & {\tt PINUDLCUNIT} & {\tt LCFITDT} \\
\hline
Standard deviation ($\sigma$) & 3.75\% & 2.31\% & 3.23\% & 1.72\% & 2.96\% & 0.99\% \\
Statistical error ($\sigma_{\rm stat}$) & \multicolumn{2}{c}{1.83\%} & \multicolumn{2}{c}{1.36\%} & \multicolumn{2}{c}{0.93\%} \\
Systematic error ($\sigma_{\rm sys}$) & 3.27\% & 1.40\% & 2.93\% & 1.05\% & 2.81\% & 0.34\% \\
\hline
40--70 keV & \multicolumn{2}{c}{10 ks}& \multicolumn{2}{c}{20 ks} & \multicolumn{2}{c}{40 ks} \\
 & {\tt PINUDLCUNIT} & {\tt LCFITDT} & {\tt PINUDLCUNIT} & {\tt LCFITDT} & {\tt PINUDLCUNIT} & {\tt LCFITDT} \\
\hline
Standard deviation ($\sigma$) & 5.53\% & 4.92\% & 4.34\% & 3.51\% & 3.39\% & 2.87\% \\
Statistical error ($\sigma_{\rm stat}$) & \multicolumn{2}{c}{4.03\%} & \multicolumn{2}{c}{3.01\%} & \multicolumn{2}{c}{2.03\%} \\
Systematic error ($\sigma_{\rm sys}$) & 3.78\% & 2.82\% & 3.12\% & 1.81\% & 2.71\% & 2.02\% \\
\hline
\multicolumn{5}{l}{$\sigma_{\rm err}=\sqrt{\sigma^2-\sigma_{\rm stat}^2}$}
\end{tabular}
\end{center}
\normalsize
\end{table}

\subsection{Comparison with the data of dark objects}\label{pinsky}

In the previous subsection, we confirmed that the background modeling is
accurate by several percents for the earth occultation data.
As described in \S\ref{bgddpin}, we consider that the background
rate somewhat depends on whether the pointing direction is toward the sky or the earth.
Although we include this effect into the model, we must check it.
We therefore assess the NXB reproducibility by utilizing the sky observation
with no apparent strong hard X-ray objects below.

\begin{figure}[htb]
\centering
\includegraphics[width=0.49\textwidth]{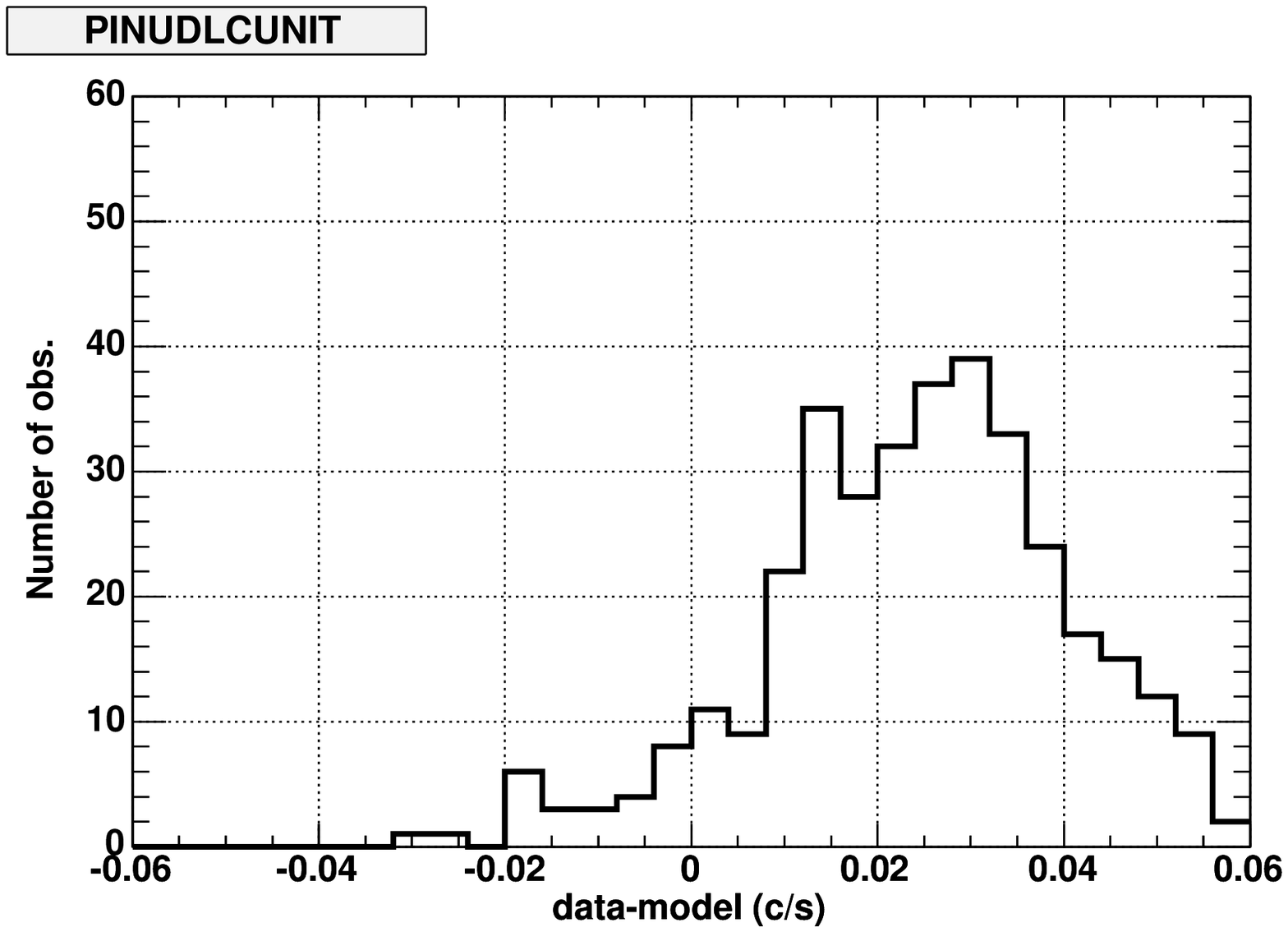}
\includegraphics[width=0.49\textwidth]{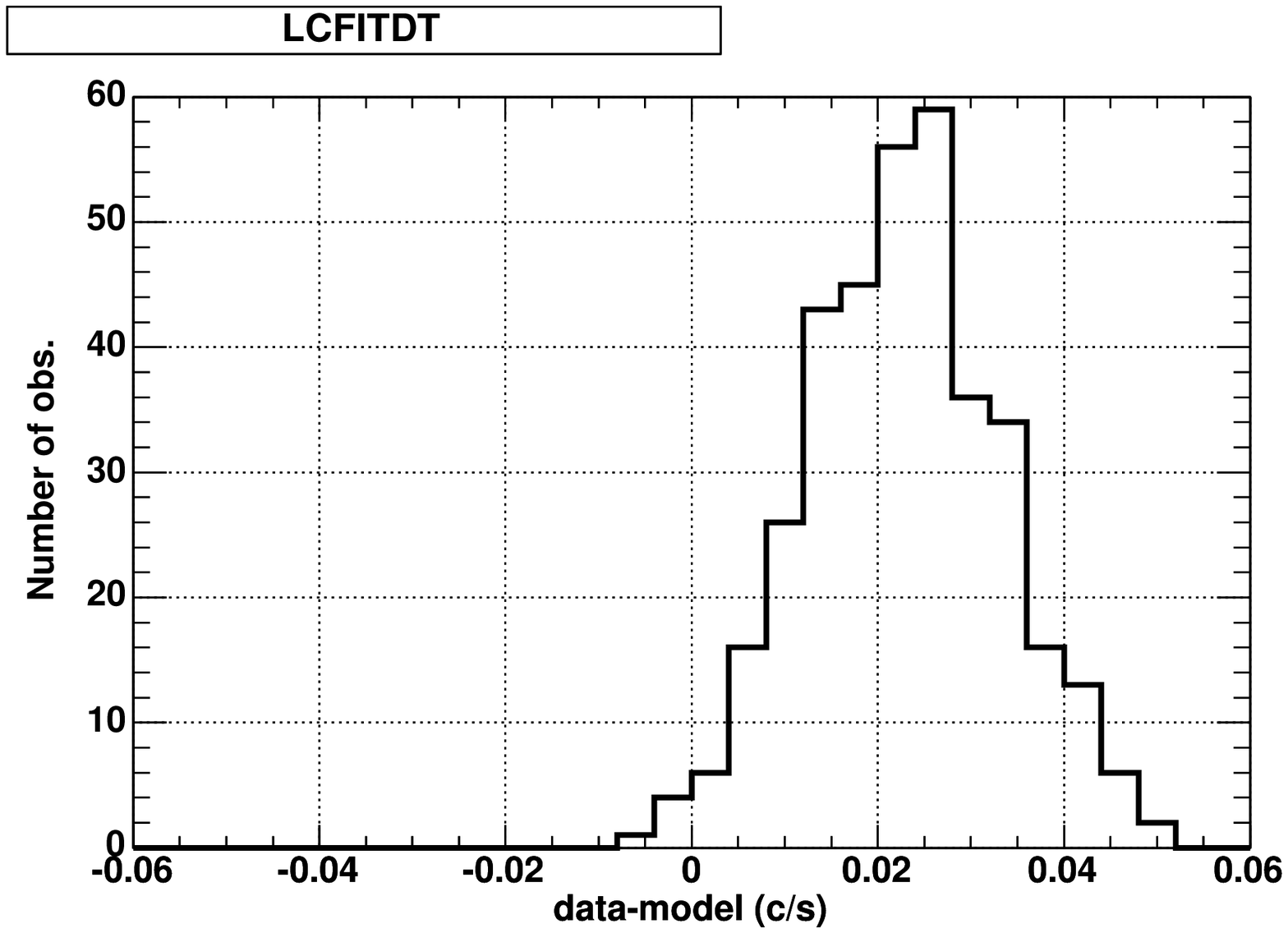}
\caption{Histogram of the residual count rate in 15--40 keV between the data 
and the NXB model for sky observations
with 10~ks integration time.
Observations with no apparent hard X-ray objects 
in XIS FOV are selected
(see text for details of the data selection).
Left and right panels show the plots for {\tt PINUDLCUNIT} and {\tt
 LCFITDT}, respectively.
}
\label{fig:IsobePlotSky}
\end{figure}

We first utilized the XIS FI data to select observations with
no strong X-ray emission in 7--12 keV
(less than 20\% above the XIS-FI NXB in entire XIS field-of-view) 
and compared the
HXD-PIN data and the NXB model count rate (10~ks exposure) in 15--40~keV, 
as shown in figure~\ref{fig:IsobePlotSky}.
Here we plotted the residual histogram in unit of count rate, not 
the ratio to the NXB model, 
since we expect a constant excess above 0 due to the CXB emission.
The standard deviation (including statistical error)
of the residual is about 5.0\% and 3.5\% of the mean NXB
count rate for {\tt PINUDLCUNIT} and {\tt
 LCFITDT}, respectively, which are somewhat larger than that 
obtained from the Earth occultation data.

One possibility is that the systematic uncertainty increases 
when observing the sky, but the source confusion limit or fluctuation of the
CXB must be considered before concluding.
The source confusion limit is estimated from the logN-logS relation 
in the hard X-ray band.
Based on the logN-logS relation obtained with the {\it INTEGRAL}
\citep{beckmann06}, the source confusion limit for the PIN field of view
becomes $8\times10^{-13}$ erg s$^{-1}$ cm$^{-2}$ (10--50 keV) at
3$\sigma$ level, at most 0.08\% of the PIN NXB.
The CXB sky-by-sky fluctuation is calculated by scaling the
HEAO-1 result of 2.8 \% with the equation $\sigma_{\rm CXB} \propto
\Omega^{-0.5} S^{0.25}$\citep{condon74}, where $\Omega$ = 15.8 deg$^2$ and $S \sim 8\times10^{-11} \ \rm erg cm^{-2} s^{-1}$ are the effective
solid angle of the observation and upper cut-off flux of the
detectable discrete sources in the field of view, respectively. In the
case of the HXD ($\Omega = 34' \times 34'$, and $S \sim 8\times
10^{-12} \ \rm erg cm^{-2} s^{-1}$), the sky-to-sky fluctuation of the CXB is
calculated to be 11\% (1$\sigma$) of CXB flux, and therefore 
may not affect the reproducibility of the NXB 
since 1$\sigma$ CXB fluctuation
corresponds to at most $\sim0.7$\% of the PIN NXB, and
then a larger standard deviation 
of the residual between data and model for the sky observation is not
fully resolved.
As a result, a standard deviation for 10 ks sky observations for the PIN
15-40 keV band becomes 2.9\%, after subtracting the contribution of
statistical error and CXB fluctuation.
This is still larger than that for the earth data.
Therefore, the reproducibility of the NXB model seems worse for sky
observations, but
one possibility is that the data selection based on the XIS could
include the source with $>0.2$ mCrab (2\% of the PIN NXB in 15--40 keV)
and such data could contribute to the standard deviation.

\begin{figure}[htb]
\centering
\includegraphics[width=0.49\textwidth]{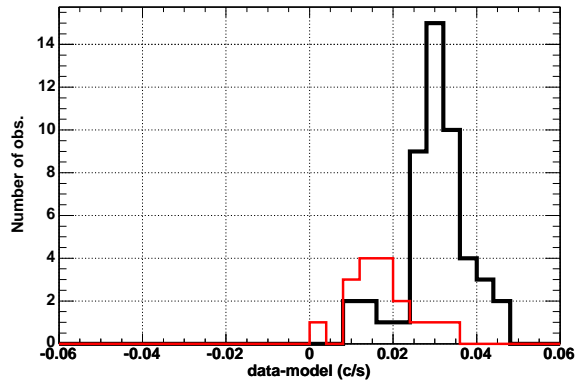}
\caption{The same as Figure~\protect\ref{fig:IsobePlotSky}, but
observations of E0102 (black) and Cygnus LOOP multi-pointing (red) are
 used and results for the background model {\tt LCFITDT} are shown.}
\label{fig:IsobePlotE0102CygLOOP}
\end{figure}

In order to avoid the possible CXB fluctuation sky by sky and source
confusion observation by observation, 
we utilized the data 
of objects which have been repeatedly observed.
Figure~\ref{fig:IsobePlotE0102CygLOOP} shows the comparison
between the sky data and the NXB model for the SNR E0102.2-7219 observations,
in which the same region of the sky is observed 
regularly for the XIS calibration purpose.
The X-ray emission from this source is thought to be constant and dark
for the HXD-PIN, but
some observations do not satisfy the selection criteria written at the
beginning of this subsection, possibly
due to sources or diffuse emission in the SMC within the XIS
field-of-view, as reported by \citet{takei08}.
Different roll angle of the field of view against the SMC may also cause
the difference of count rate among observations.
Yet, we used all E0102.2-7219 observations in order to compare
sky data and the NXB model of as many observations as possible.
We also plotted the data and the NXB model 
of the Cygnus Loop multi-pointing observations in red.
These observations were performed for the XIS calibration and cover the 
regions within 1.5 degree in radius, and thus the CXB
fluctuation within the field of view of the HXD-PIN can be reduced.

We see a clear difference of the residual between two sets of observations.
The difference of residuals between two data sets
by 0.01--0.015~${\rm c~s^{-1}}$
could be due to a weak hard X-ray 
emission inside the PIN field-of-view of E0102.2-7219 observations.
We look at a Swift/BAT archival image
\footnote{\tt http://skyview.gsfc.nasa.gov/ }, 
but no significant sources with 0.3
mCrab (0.02 c s$^{-1}$ for PIN) are found.
Since this region is in the SMC, many candidates of hard X-ray sources
exist, and a sum of their flux would explain the above excess flux.

From figure \ref{fig:IsobePlotE0102CygLOOP}, 
the residual for E0102.2-7219 data concentrates around 0.03 c~s$^{-1}$
with a narrower width than that for all the dark objects 
in figure~\ref{fig:IsobePlotSky}, 
but a few observations show an extraordinary small residual below 0.01
c~s$^{-1}$.
A standard deviation of the residual, including statistical errors, 
is 0.0080 c~s$^{-1}$ and 0.0077 c~s$^{-1}$ (2.7\% and 2.6\% 
of the mean NXB rate) for E0102.2-7219 and Cygnus Loop, respectively.
This is smaller than that obtained from many dark observations, but
somewhat larger than that for the earth observations.
Three data points below 0.015 c~s$^{-1}$ for E0102.2-7219 are from 
observation on Feb. 10--11, 2007, and thus NXB accuracy is not good
around this observation.
As noted in \S\ref{noearth}, it was found that the NXB model accuracy is
not good for specific periods.
In addition, four Cygnus Loop observations were performed in May 2006, when
the background accuracy is relatively worse because the HXD operation mode
was changed several times (\S2.2).
When excluding these data, the standard deviation becomes 2.3\% and
2.1\% of the mean NXB rate for E0102.2-7219 and Cygnus Loop,
respectively, and these values are almost similar to that for the earth
observations.
Therefore, the PIN NXB background accuracy is almost the same between
earth and sky observations in most cases.
We return to this issue in \S\ref{noearth}.

\begin{figure}[htb]
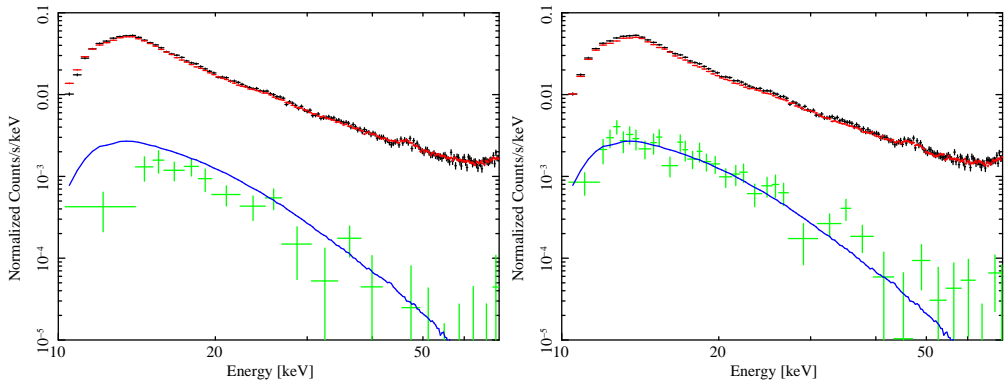

\includegraphics[width=0.3\textwidth,angle=-90]{fig14a.eps}
\includegraphics[width=0.3\textwidth,angle=-90]{fig14b.eps}
\caption{Averaged spectra of sky observations (black) and the NXB model (red) of
Cygnus LOOP multi-pointing observations, together with the subtracted spectrum
(green) and the CXB model (blue) by Boldt (1987).
Top and bottom panels show the plots for {\tt PINUDLCUNIT} and {\tt
 LCFITDT}, respectively.
}
\label{fig:CXBestimate}
\end{figure}

Next, we investigate the accuracy of background-subtracted spectra by
referring to the CXB.
If we assume that the Cygnus Loop regions are free from any weak hard X-ray emission,
we can regard the averaged spectrum as a sum of the CXB plus the NXB.
The NXB-model-subtracted spectrum is compared with the 
spectrum of the CXB model
by \citet{boldt87} in figure~\ref{fig:CXBestimate}.
We see a very good agreement between the subtracted spectrum (green crosses)
and the CXB model (blue ones) within 1\% error.


\begin{figure}[htb]
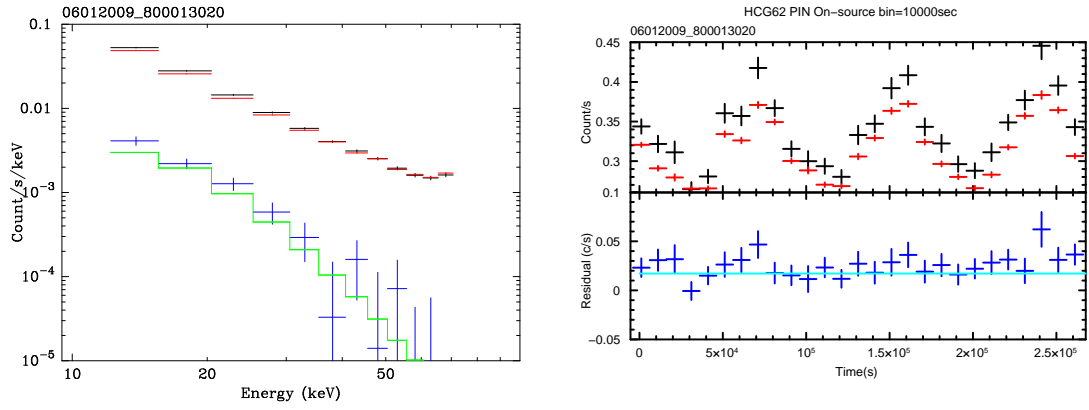

\centering

\includegraphics[angle=-90,width=0.4\textwidth]{fig15a.eps}
\hspace{5mm}
\includegraphics[angle=-90,width=0.4\textwidth]{fig15b.eps}

\caption{
Left panels are comparison of spectra between the data (black) and BGD
 model {\tt LCFITDT} (red) 
for observation of HCG 62, which is thought to contain
no known strong hard X-rays. 
Blue cross data points are residuals after the BGD model is
subtracted, and
the green solid line indicates
a typical CXB spectrum.
Right figure is the comparison of the light curve in 15--40~keV
between the data (black) and BGD model {\tt LCFITDT} (red) 
for observation of HCG 62.
The bottom panel shows the residuals. A time bin is 10000 sec.
The horizontal line in the bottom panel indicates the
CXB level.
}
\label{fig:blankSkyPIN}
\end{figure}

In order to check the NXB reproducibility in a single observation,
we also compared
the spectrum and the light curve of objects
whose signal is expected to be negligible for the HXD-PIN.
Example of comparison of spectra 
are shown in Figure~\ref{fig:blankSkyPIN} left. 
The background-subtracted spectrum and the CXB model by Boldt (1987)
is given in blue and green histograms, respectively.
No systematic difference is seen between
them up to 60~keV.
Figure~\ref{fig:blankSkyPIN} right also 
compared the light curves between the data and the NXB model 
in 15--40~keV band with a time bin of 10~ks.
Residuals are consistent with the CXB level in 
${\rm \pm \sim 0.02~c~s^{-1}}$, or $\sim 7\%$
of the total NXB count rate.

\clearpage
\section{Reproducibility of the GSO NXB}

\subsection{Data Reduction}

The data reduction is almost the same as the check of the PIN NXB,
but we discard the period when the total GSO count rate is less
than 15 cts sec$^{-1}$.
Such a low count rate occurs when only 1/4 of GSO data are output to the
telemetry for reducing the data size, usually during the data rate L
(usually during SAA, low COR period, or earth occultation).
However, this 1/4 mode sometimes continues 
even in the regular observation period
\footnote{Data rate is medium or high or superhigh}, 
and becomes contained in the clean event.
The background model does not support such a mode.

\subsection{Comparison with the Earth Occultation Data}

\begin{figure}[htb]
\centering
\includegraphics[width=0.42\textwidth]{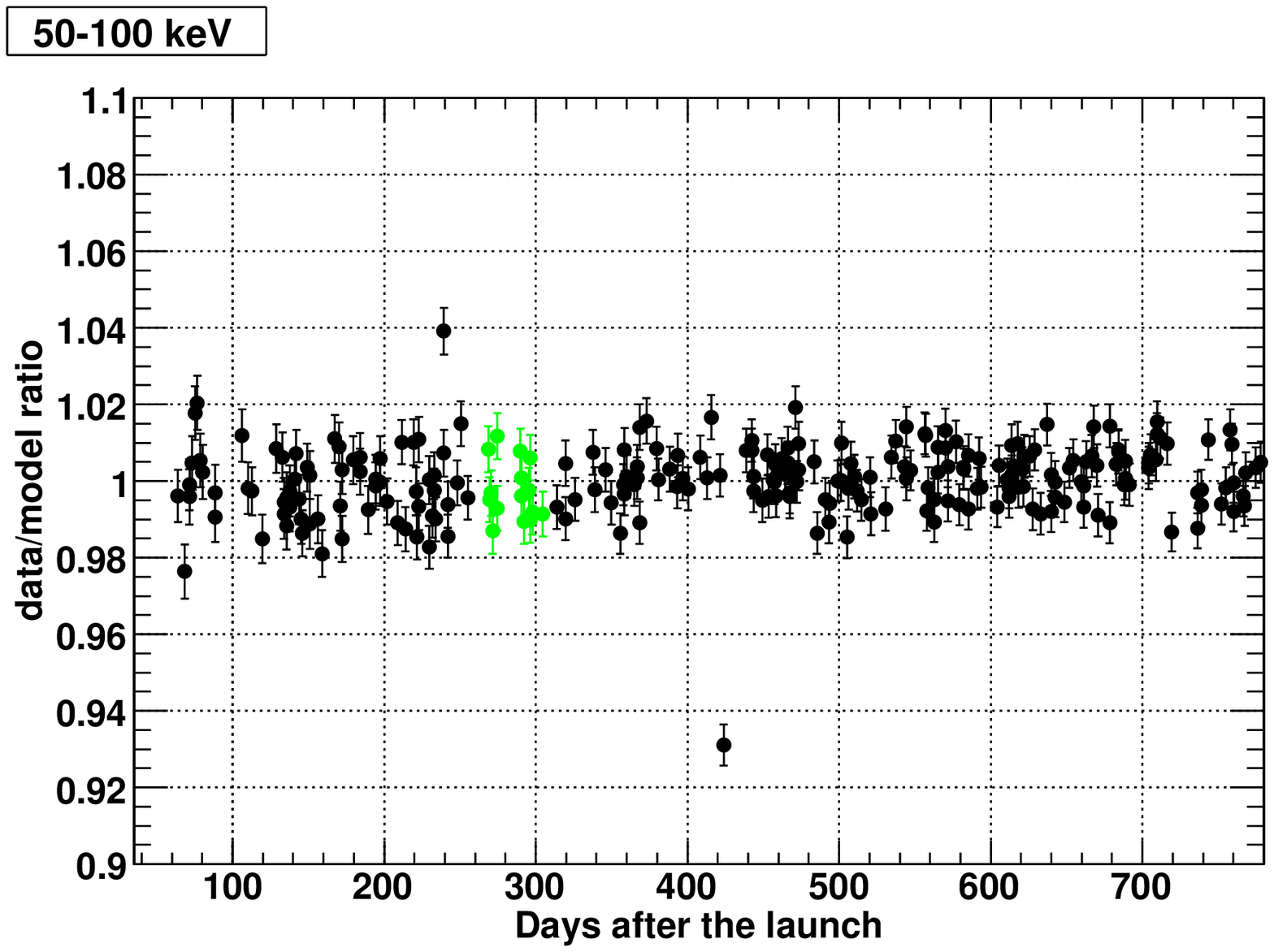}
\includegraphics[width=0.42\textwidth]{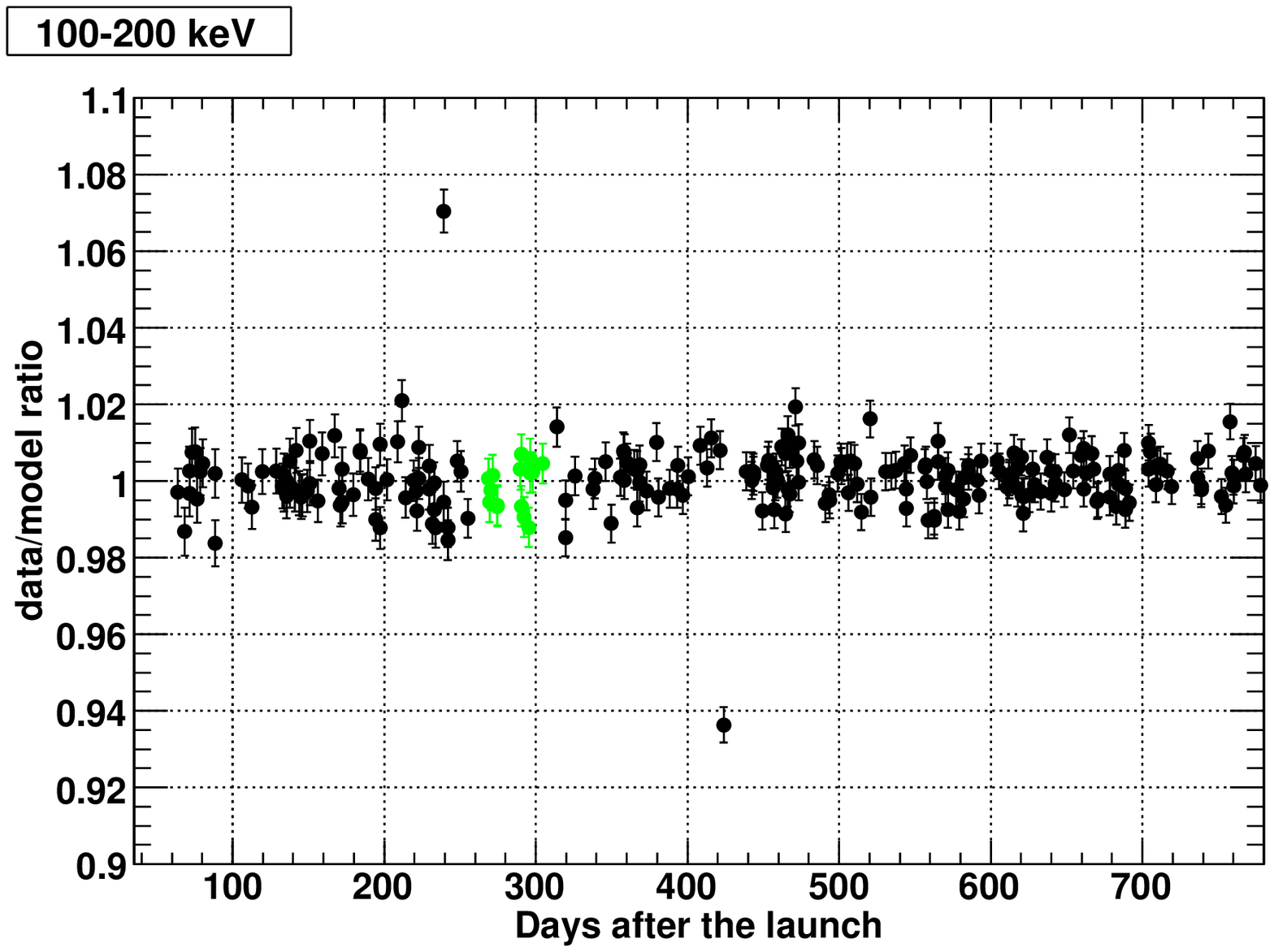}
\caption{
(left) A comparison of the NXB count rate in the earth occultation
in 50--100~keV between the data and the model prediction.
Data from 2006 March 23 to May 13 are shown by green crosses (see text).
(right) The same plot but for 100--200~keV.
}
\label{fig:IsobePlotSWG}
\vspace*{0.5cm}
\centering
\includegraphics[width=0.42\textwidth]{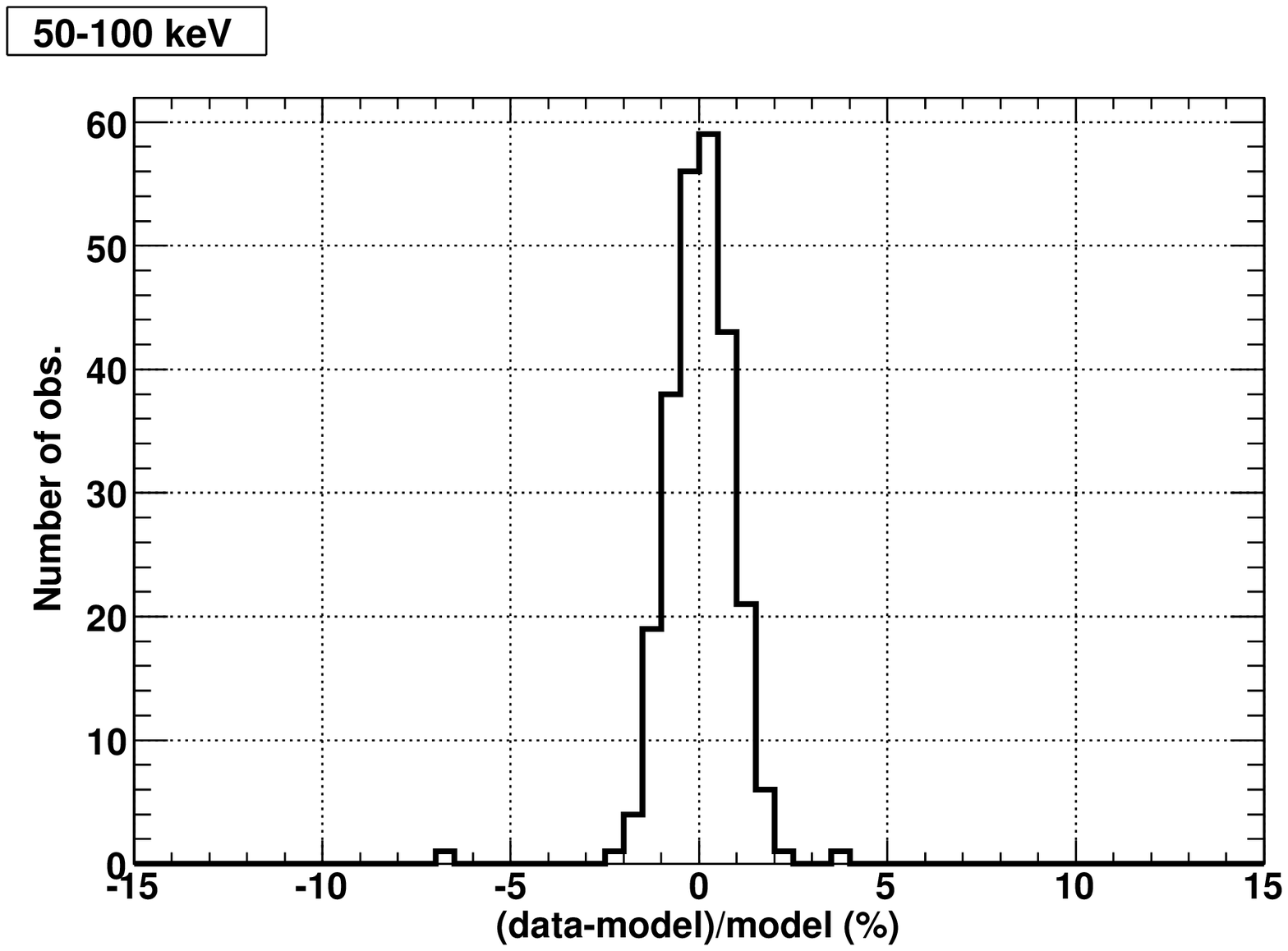}
\includegraphics[width=0.42\textwidth]{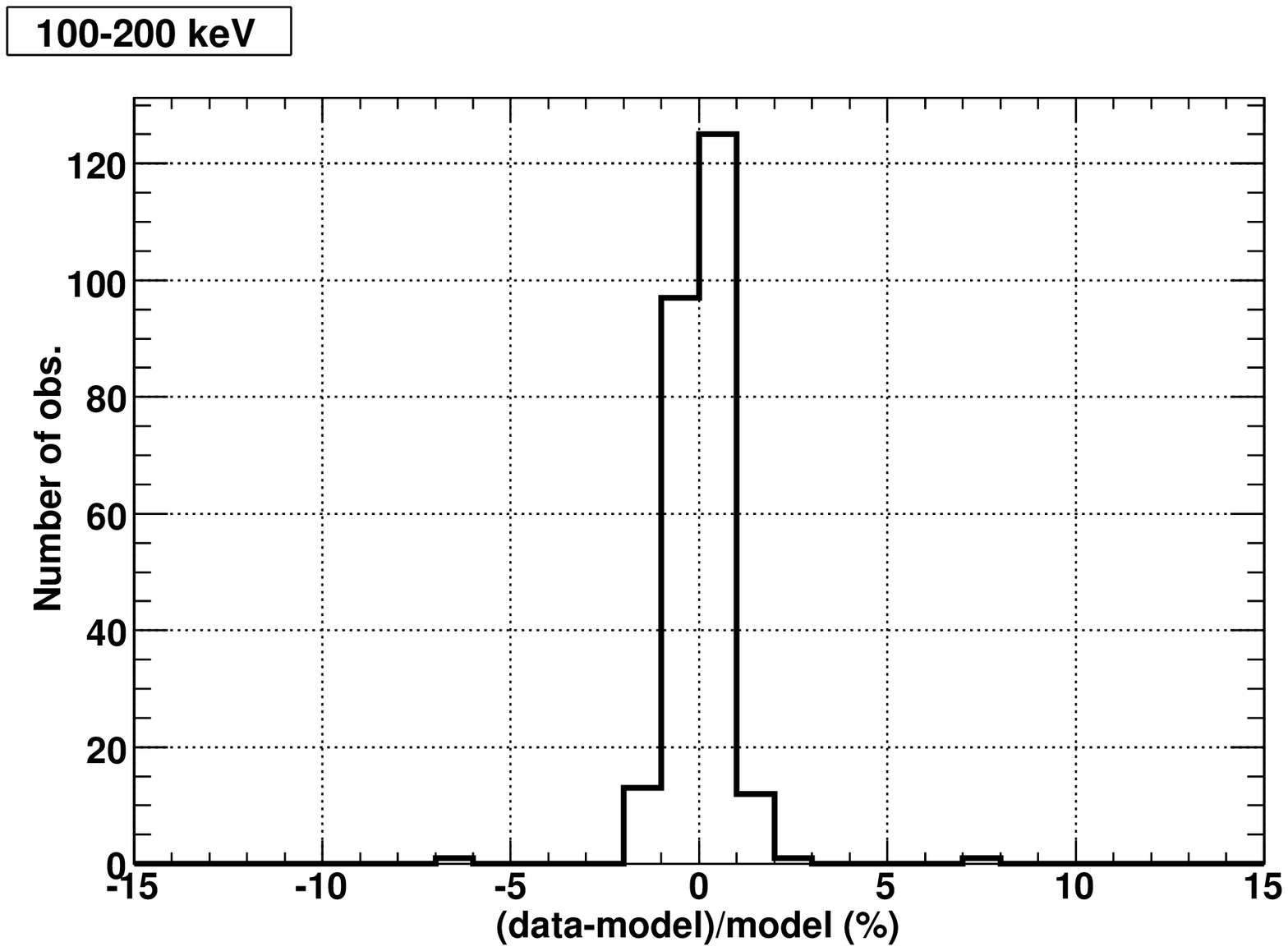}
\caption{
(left) Distribution of fractional residual of the NXB count rate between
the data and the model prediction in the 50--100 keV band.
(right) The same plot but for the 100--200~keV band.
}
\label{fig:IsobePlotDist}
\end{figure}

Each observation is divided into several periods during which the
exposure is 10 ks.
When the exposure of earth occultation within one observation is less 
than 10 ks, we do not use the data.
Accordingly, the GSO count in each data set is typically $(4-9)\times10^4$, and
thus the statistical error is typically 0.3--0.5 \%.
In Figure~\ref{fig:IsobePlotSWG}, 
we compared the NXB count rates in each data set 
between the earth occultation data
and the background model in 50--100~keV and 100--200~keV range, 
against the elapsed day after the Suzaku launch.

Most of the data and model agree within $\sim 2~\%$ with no time dependence,
and no significant difference in reproducibility is
seen between models even in the period of 2006 March 23 to May 13
during which the GSO LD level was changed.
There are some data points, which significantly deviate from 1.0 
by $>$2\%, corresponding to observations during which the HXD
observation mode is not nominal and the background model cannot be applied.
Figure \ref{fig:IsobePlotDist} plots the distribution of the NXB 
count rate ratio between the data and the model prediction. 
In both energy band, the distribution is well represented by a
Gaussian with $\sigma$ of 0.72 \% and 0.59 \% in the 50--100 keV and
100--200 keV band, respectively.
Considering the average of statistical error of 0.40 \% and 0.36 \%, the
1$\sigma$ systematic error is estimated to be 0.60 \% and 0.47 \% 
in the 50--100 keV and
100--200 keV band, respectively.
Note that the systematic error depends on the event selection criteria
(e.g. COR and energy band) and the integration time.
Table \ref{gsosigma} summarizes a standard deviation in each energy
band as a systematic error.
There is an energy band dependency, but the systematic error is less
than 1\% at 1$\sigma$ level in any energy band.

\begin{figure}[htb]
\vspace*{-3cm}
\centerline{\includegraphics[angle=0,width=0.5\textwidth]{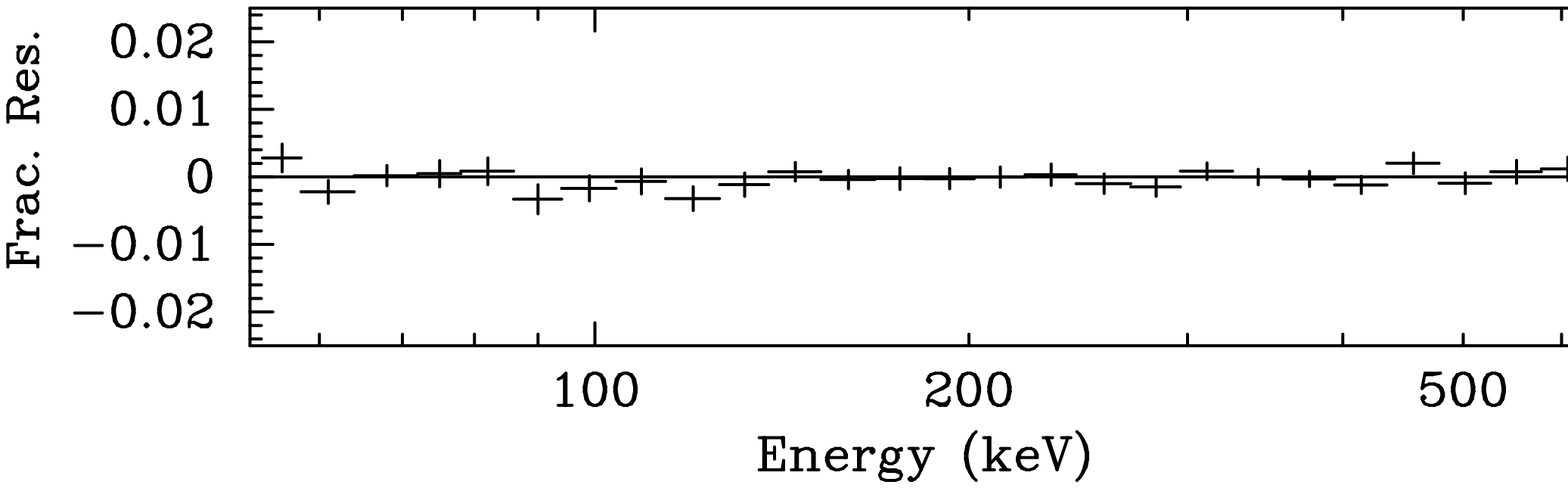}}
\vspace*{-3cm}
\centerline{\includegraphics[angle=0,width=0.5\textwidth]{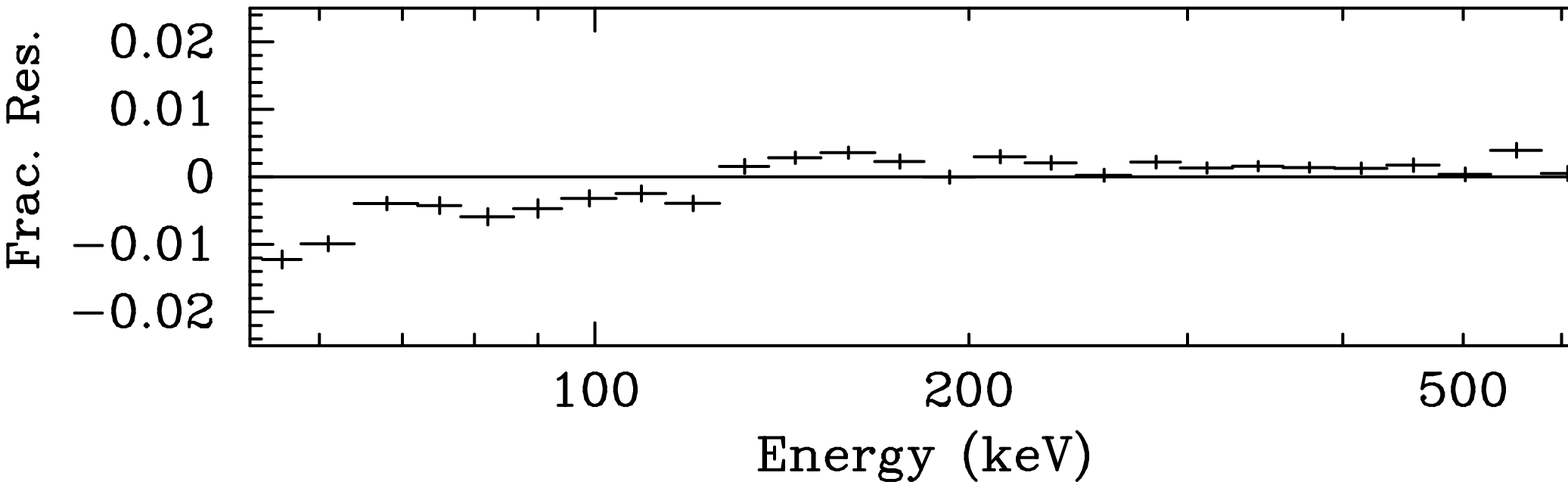}}
\caption{Fractional residual of the spectra after subtracting the GSO
background model from the earth data (top) and the sky data (bottom),
summed over 88 observations of dark objects.}
\label{fig:EarthSkySpec}
\end{figure}

Figure \ref{fig:EarthSkySpec} left shows the 
comparison of the spectra for earth data and background model,
summed over 88 observations of dark objects.
The total exposure is 923 ks.
It can be seen that the data and model well agree with each other in all
the energy band within 1 \%.

\begin{table}[ht]
\caption{Standard deviation (1$\sigma$) 
of residuals between the GSO earth data and
 model in each energy band.}
\label{gsosigma}
\begin{center}
\small
\begin{tabular}{ccccc}
\hline\hline
 & 50--100 keV & 100--200 keV & 200--300 keV & 300--500 keV \\
\hline
Standard deviation ($\sigma$) & 0.72\% & 0.59\% & 0.87\% & 0.71\% \\
Statistical error ($\sigma_{\rm stat}$) & 0.40\% & 0.36\% & 0.46\% & 0.34\% \\
Systematic error ($\sigma_{\rm sys}$) & 0.60\% & 0.47\% & 0.74\% & 0.63\% \\
\hline
\multicolumn{5}{l}{$\sigma_{\rm err}=\sqrt{\sigma^2-\sigma_{\rm stat}^2}$}
\end{tabular}
\end{center}
\normalsize
\end{table}

\subsection{Comparison with the Data of Dark Objects}
\begin{figure}[htb]
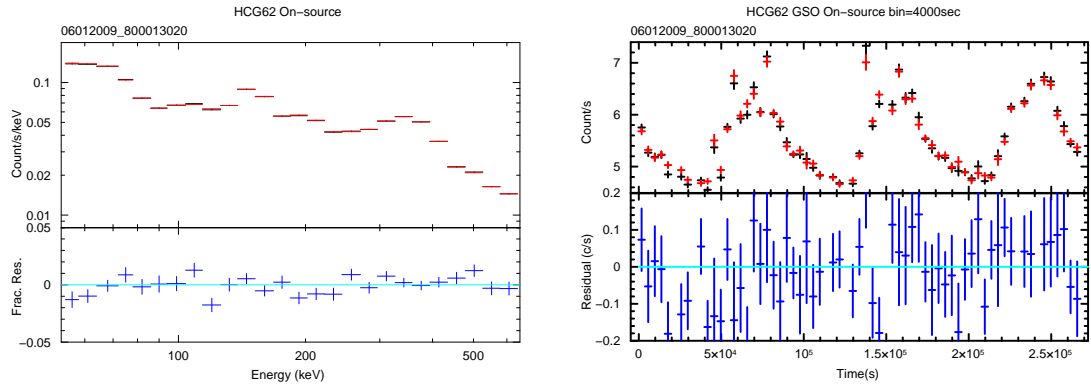

\centering

\includegraphics[angle=-90,width=0.4\textwidth]{fig19a.eps}
\hspace{5mm}
\includegraphics[angle=-90,width=0.4\textwidth]{fig19b.eps}

\caption{
The left figure is a comparison of spectra between the data (black) 
and BGD model (red) 
for observation of HCG 62 whose sky region
contains no known strong hard X-rays. Fractional residuals are given by blue
crosses in the bottom panel.
Right figure is a comparison of the light curve in 50--100~keV
between the data (black) and BGD model {\tt LCFITDT} (red) 
for observation of HCG 62.
The upper panel shows the light curve and the
lower panel shows the residual. A time bin is 4000 sec.
}
\label{fig:blankSky}
\end{figure}

In this section, we compared 
the NXB model with the sky data of dark objects, 
whose signal is expected to be negligible for the HXD-GSO. 
Example of comparison of spectra for dark objects
are summarized in Figure~\ref{fig:blankSky} left. 
Unlike the PIN, the CXB is negligible in the GSO band.
No systematic difference is seen between
the data and BGD model spectra, indicating that the
background model is applicable for sky observations.
We also compared the data and the NXB model light curves 
as shown in figure~\ref{fig:blankSky} right.
for the 50--100~keV band with a time bin of 4000 sec.
Residuals are mostly within 2 \% of the total count rate, and
we see some
modulations of a peak-to-peak amplitude up to $\sim 0.1~{\rm c~s^{-1}}$
in a cycle of $\sim$ 1 day.

Figure \ref{fig:EarthSkySpec} right shows the 
comparison of the spectra for on-source data and background model,
summed over 88 observations of dark objects.
The exposure is 2430 ks.
Although systematic difference is seen between the
data and the model, they agree with each other within
1\% in the whole energy band.

\clearpage
\section{Cause of Current Systematic Errors of the BGD model}\label{noearth}

\begin{figure}[htb]
\vspace*{-3cm}
\includegraphics[width=0.5\textwidth]{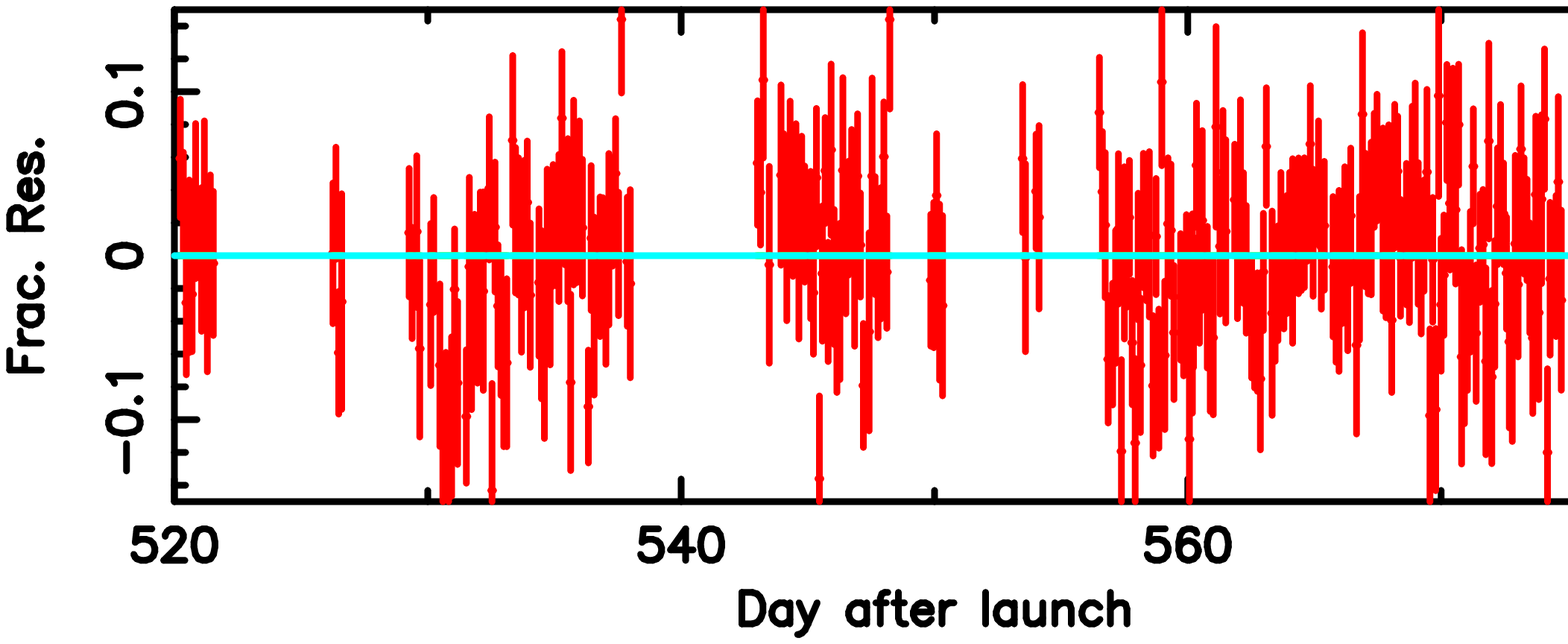}
\includegraphics[width=0.5\textwidth]{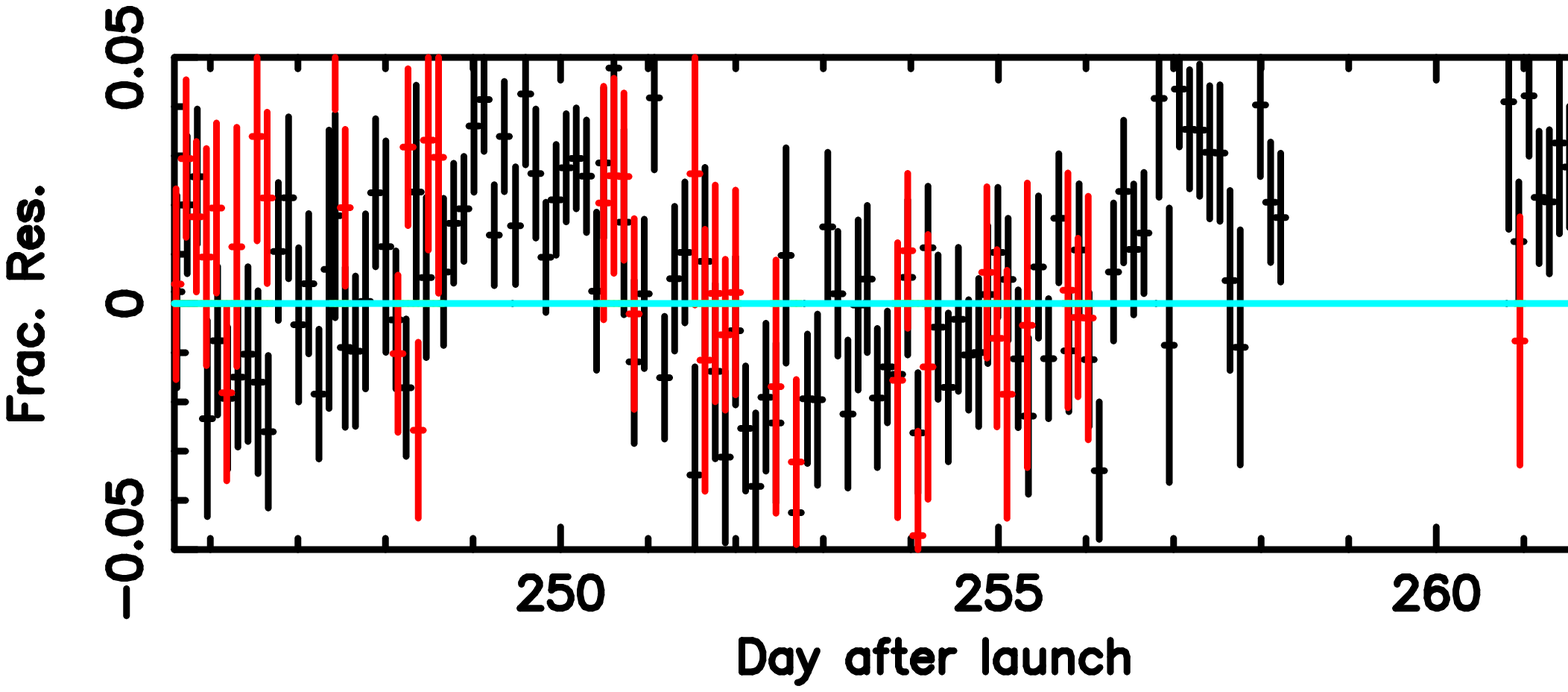}
\caption{Light curve of fractional residual of count rate after
 subtracting the background model in
 15--40 keV and 70--100 keV for the PIN (left) and GSO (right), respectively. 
Black and red represent
 the residual of on-source data and earth data, respectively. 
}
\label{fig:longlc}
\end{figure}

Although we have developed the background model as accurately as
possible, it does not completely reproduce the NXB.
This may cause the issues described in \S\ref{pinsky}; background
reproducibility is not good for some specific observations.
For studying the BGD systematic error related with this issue,  
it is very useful to investigate the background-subtracted light curve 
of the count rate in several tend days.
Figure \ref{fig:longlc} shows light curves of fractional residual of count
rate after subtracting the background model in 15--40 keV and 70--100
keV for PIN and GSO, respectively. 
Data of both on-source and earth-occultation are plotted as different
colors for the GSO data.
We do not show the on-source data for the PIN, since the PIN data contain the
CXB signal and also often the target signal, and they are not useful for
studying the BGD systematics.
Together with the light curve of a shorter time scale in figure
\ref{fig:blankSky}, it can be seen that the BGD systematic
uncertainty is composed of two components; one with a shorter time scale
within 1 day and another with a longer time scale of $>$1 day.
The latter uncertainty appears as a modulation with a time scale of 
several days in the residual light curve, and its behavior is common 
between on-source and earth data. 
This behavior is clearly seen for the GSO. 
For the PIN, the modulation is a less clear but significant modulation 
at several percents exists.
This trend is also found for the PIN NXB-subtracted light curve 
around the E0102.2-7219 observation on Feb. 10--11, 2007, 
leading to somewhat worse reproducibility as described in \S\ref{pinsky}.

Accordingly, by studying the light curve as above, 
the background level is more accurately determined
than by simply subtracting the background model.
As can be seen from the trend in figure \ref{fig:longlc}, the
reproducibility is generally worse around the observation during which
earth occultation data are not available for $>$1 days.
This is because the NXB modeling refers to the earth data.
This trend could be smaller when sky observations with no significant
PIN/GSO signal are utilized for the NXB modeling.
Since celestial GSO signal is negligible in most of observations, 
the choice of sky observations is not difficult for the GSO NXB
modeling.
On the other hand, this is not the case for the PIN, the choice of sky
observations must be paid much attention to.

\begin{figure}[htb]
\centering
\includegraphics[angle=0,width=0.4\textwidth]{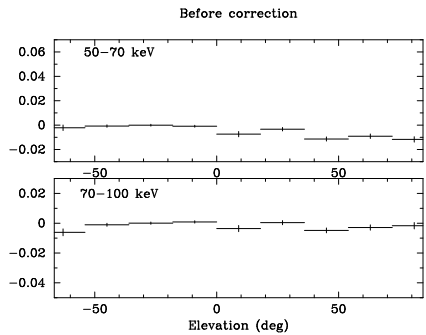}
\caption{
The residual of GSO background subtraction against the earth elevation angle.
}
\label{elev2}
\end{figure}

Apart from the above issue, for the GSO NXB, the elevation
dependence of $\sim$1 \% has been found below 70 keV as shown in
figure \ref{elev2}.
Although the dependence is very small, we will include such dependence
into the GSO NXB model as the PIN NXB model in the near future.
This will improve the systematic negative residual in the low energy
band in figure \ref{fig:EarthSkySpec}.

\section{Current Sensitivity of the HXB}

\begin{figure}[htb]
\centering
\includegraphics[angle=90,width=0.6\textwidth]{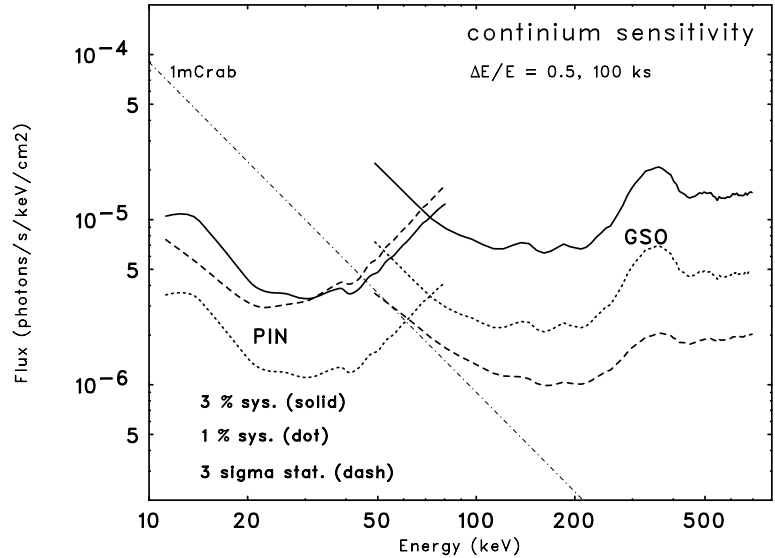}
\caption{Continuum sensitivity of the HXD. The solid and dot lines
 represent the sensitivity limited by the systematic error of the NXB
 modeling at 3\% and 1\% accuracy, respectively. The dashed line
 represents the sensitivity limited by the statistical error for 100 ks observation.
}
\label{sensitivity}
\end{figure}

As described in the previous sections, the current HXD NXB modeling
achieves 1--3\% reproducibility, but this accuracy depends on the
integration time, energy band, period, and so on.
Generally, a shorter integration time or a narrower energy band introduce a
larger statistical error, and thus the accuracy of the NXB subtraction
is limited by the photon statistics of the background count rate.
Figure \ref{sensitivity} shows a comparison of the sensitivity limited
by the systematic and statistical errors.
Since the PIN-NXB count rate is very low, the statistical error becomes
dominant for a shorter integration time.
On the contrary, the GSO sensitivity is almost determined by the
systematic error.
Note that this figure represents the sensitivity in a wide energy band
$\Delta E$ of $0.5E$ at a given energy $E$ for a 100 ks exposure.
In the spectral analysis, the spectral bin size often corresponds to
narrower energy band, and thus the statistical error in each spectral
bin is larger.
Furthermore, the PIN sensitivity is also dependent on the CXB
sky-to-sky fluctuation, which is similar level to the current PIN-NXB
reproducibility.

\section{Summary}

We modeled the HXD PIN/GSO NXB by utilizing several parameters,
including particle monitor counts and satellite orbital/attitude information.
Current reproducibility of the NXB model 
is estimated to be less than 3\% 
(PIN 15--40 keV) and 1\% (GSO 50--100 keV) for 
more than 10 ksec exposure.
The reproducibility is generally worse around the observation during which
no earth occultation data are contained.

The authors wish to thank all members of the Suzaku Science Working
Group, for their contributions to the instrument preparation, spacecraft
operation, software developement, and in-orbit calibration. This work is
partly supported by Grants-in-Aid for Scientific Research by the
Ministry of Education, Culture, Sports, Science and Technology of Japan 
(1479206 and 1479201).


\end{document}